\documentclass{aa}

\usepackage[varg]{txfonts}
\usepackage{graphicx}
\usepackage{natbib}
\bibpunct{(}{)}{;}{a}{}{,}   

\usepackage{multirow}        
\usepackage{dcolumn}         

\begin{document}

\title{The ALMA Band~9 receiver}
\subtitle{Design, construction, characterization, and first light}

\author{A.M.~Baryshev \inst{1,2}
   \and R.~Hesper \inst{1}
   \and F.P.~Mena \inst{2,3}
   \and T.M.~Klapwijk \inst{4}
   \and T.A.~van Kempen \inst{5,6}
   \and M.R.~Hogerheijde \inst{5}
   \and B.D.~Jackson \inst{2}
   \and J.~Adema \inst{1}
   \and G.J.~Gerlofsma \inst{1}
   \and M.E.~Bekema \inst{1}
   \and J.~Barkhof \inst{1}
   \and L.H.R.~de Haan-Stijkel \inst{1}
   \and M.~van den Bemt \inst{1}
   \and A.~Koops \inst{1}
   \and K.~Keizer \inst{2}
   \and C.~Pieters \inst{2}
   \and J.~Koops van het Jagt \inst{1}
   \and H.H.A.~Schaeffer \inst{2}
   \and T.~Zijlstra \inst{4}
   \and M.~Kroug \inst{4}
   \and C.F.J.~Lodewijk \inst{4}
   \and K.~Wielinga \inst{7}
   \and W.~Boland \inst{5,8}
   \and M.W.M.~de Graauw \inst{6}
   \and E.F.~van Dishoeck \inst{5}
   \and H.~Jager \inst{2,6}
   \and W.~Wild \inst{1,2,9}
}

\institute{Kapteyn Astronomical Institute, University of Groningen, Postbus~800, 9700~AV Groningen, The Netherlands\\     
           \email{r.hesper@astro.rug.nl}
      \and SRON Netherlands Institute for Space Research, Postbus~800, 9700~AV Groningen, The Netherlands\\               
           \email{a.m.baryshev@sron.nl}
      \and Department of Electrical Engineering, Universidad de Chile, Av. Tupper~2007, Santiago, Chile\\                 
           \email{pmena@ing.uchile.cl}
      \and Kavli Institute of Nanoscience, Delft University of Technology, Lorentzweg~1, 2628~CJ Delft, The Netherlands\\ 
           \email{t.m.klapwijk@tudelft.nl}
      \and Leiden Observatory, Leiden University, PO Box 9500, 2300~RA, Leiden, The Netherlands\\                         
           \email{kempen@strw.leidenuniv.nl}
      \and Joint ALMA Offices, Av. Alonso de C\'ordova~3107, Vitacura - Santiago, Chile                                   
      \and Mecon Engineering BV, Koopmanslaan~25, 7005~BK Doetinchem, The Netherlands                                     
      \and NOVA, J.H.~Oort Building, P.O. Box~9513, 2300~RA Leiden, The Netherlands                                       
      \and ESO, Karl-Schwarzschild-Stra\ss{}e~2, 85748 Garching, Germany                                                  
}

\date{Received 17 December 2014 / Accepted 24 February 2015}

\abstract
   {}
   {We describe the design, construction, and characterization of the Band~9
    heterodyne receivers (600--720\,GHz) for the Atacama Large
    Millimeter/submillimeter Array (ALMA). First-light Band~9 data, obtained
    during ALMA commissioning and science verification phases, are presented as
    well.}
   {The ALMA Band~9 receiver units (so-called ``cartridges''), which are
    installed in the telescope's front end, have been designed to detect and
    down-convert two orthogonal linear polarization components of the light
    collected by the ALMA antennas. The light entering the front end is
    refocused with a compact arrangement of mirrors, which is fully contained
    within the cartridge. The arrangement contains a grid to separate the
    polarizations and two beam splitters to combine each resulting beam with a
    local oscillator signal. The combined beams are fed into independent
    double-sideband mixers, each with a corrugated feedhorn coupling the
    radiation by way of a waveguide with backshort cavity into an
    impedance-tuned superconductor-insulator-superconductor (SIS) junction that
    performs the heterodyne down-conversion. Finally, the generated
    intermediate frequency (IF) signals are amplified by cryogenic and
    room-temperature HEMT amplifiers and exported to the telescope's IF back
    end for further processing and, finally, correlation.}
   {The receivers have been constructed and tested in the laboratory and they
    show an excellent performance, complying with ALMA requirements.
    Performance statistics on all 73 Band~9 receivers are reported.
    Importantly, two different tunnel-barrier technologies (necessitating
    different tuning circuits) for the SIS junctions have been used, namely
    conventional AlO$_x$ barriers and the more recent high-current-density AlN
    barriers. On-sky characterization and tests of the performance of the
    Band~9 cartridges are presented using commissioning data. Continuum and
    line images of the low-mass protobinary IRAS~16293-2422 are presented which
    were obtained as part of the ALMA science verification program. An 8\,GHz
    wide Band~9 spectrum extracted over a 0.3"$\times$0.3" region near
    source~B, containing more than 100 emission lines, illustrates the quality
    of the data.}
   {}


\maketitle


\section{Introduction}

The Atacama Large Millimeter/submillimeter Array (ALMA) is the largest
millimeter and sub-millimeter astronomical facility ever built
\citep{wooten-ieee}. When fully operational, it will consist of 66 movable
antennas (54 with a 12-meter dish and 12 with a 7-meter dish). Every antenna of
the array holds a cryostat (front end) located near the secondary focus with
room for ten different dual-polarization heterodyne receivers to cover the
atmospheric transmission windows between 31\,GHz and 950\,GHz. At the time of
writing, ALMA is already operational at about 75 percent of its full capacity,
using over 50 antennas. The first generation of front ends delivered to the
ALMA site are populated with receivers for Band~3 (89--116\,GHz), Band~6
(211--275\,GHz), Band~7 (275--370\,GHz), and Band~9 (602--720\,GHz). Other
receivers are either close to full installation at the observatory (Bands~4, 8,
and~10), in construction (Band~5), or under development (Bands~1 and ~2).

Physically, the receivers are built as independent units called ``cartridges'',
presenting a uniform mechanical, thermal, and electronic interface to the
telescope's front end. Since many components needed for such high-performance
receivers are not available off the shelf, the cartridge concept enabled the
assembly of independent collaborative teams consisting of instrument scientists
and experts on superconducting devices, extremely low-noise amplifiers, and
local oscillators, making them jointly responsible for optimization and
delivery of a particular frequency band. For instance, each assigned
atmospheric transmission band leads to different solutions for the
superconducting tunnel devices used in the mixers of all but the two lowest
bands. For Band~9, the superconducting technology was developed in close
collaboration between the Kavli Institute of Nanoscience at Delft and the NOVA
group at the University of Groningen, where the latter also carried out the
rest of the cartridge design, assembly, and radio frequency (RF) testing. The
cartridge concept has proven to be very practical for rapid upgrading and
maintenance, and it is now also used outside of the ALMA context
\citep{sugimoto-pasj}.

The receivers for the different ALMA bands are spatially distributed within the
focal plane \citep{carter2004optics} so that the telescope needs to be
repointed when another frequency band is selected. Owing to the physical size
of the telescope beam, Bands~1--4 have reflective optical elements external to
the cryogenic front end. For Bands~5--10 all coupling and polarization
splitting optics are located within the cryostat. A water vapor radiometer is
located outside the front end to measure the atmospheric water vapor content
above each of the ALMA antennas for individual phase correction
\citep{Nikolic13}. A robotic arm with a two-temperature calibrator device,
solar filter, and quarter-wave plate allows for proper calibration of the
receivers during operation.

In a heterodyne receiver, the radio frequency (RF) signal to be studied, at
angular frequency $\omega_{RF}$, is mixed with a well-determined local
oscillator (LO) signal at $\omega_{LO}$, to be down-converted to an
intermediate frequency (IF) with $\omega_{IF} = \left|\omega_{RF}-\omega_{LO}
\right|$. The IF signal contains all the information of the original RF signal
but at a much lower frequency, which can be further processed electronically.
In the simplest, so-called double-sideband (DSB) configuration, a heterodyne
receiver cannot distinguish whether the resulting IF signal corresponds to an
RF signal that has a higher or a lower frequency than $\omega_{LO}$. In other
words, in the IF band, the down-converted lower side band (LSB) and upper side
band (USB) are superimposed. This results in extra noise on the IF output if
the observed RF signal is present in only one sideband as in the case of
spectral observations. In contrast, a sideband-separating (2SB) heterodyne
receiver has two separate IF outputs, one for RF signals down-converted from
the LSB and one for those down-converted from the USB. This is achieved at the
expense of extra complexity in the RF and IF circuitry, making it considerably
more difficult to implement at higher frequencies. As a consequence, in the
original ALMA design, Bands~3, 6, and~7 were chosen to have 2SB mixers, while
Band~9 opted for DSB mixers. Having only one IF channel per polarization, the
IF system of Band~9 was designed to cover a range of 4--12\,GHz instead of
4--8\,GHz as on the 2SB bands.

Improvements in micromachining techniques over the past few years now allows
fabrication of 2SB receivers for these high frequencies as well
\citep{mena-mtt}, and so they can be considered for future upgrading of the
Band~9 receivers \citep{hesper09, khudchenko11, khudchenko12}.

In this paper we present the technical details and performance of the full
series of Band~9 receivers. In Sect.~\ref{sec:dc} we present an overview of the
receiver concept and the design and construction of its components. In
Sect.~\ref{sec:lab} the laboratory performance of the assembled receiver is
presented, with particular attention to the beam pattern, noise temperature and
gain compression results. Statistics on the measured performance of the full
series of ALMA Band~9 receivers is also reported. In Sect.~\ref{sec:science},
we present the performance of the receiver in a single ALMA antenna as well as
within the larger array. Several data sets taken during the commissioning time
of ALMA illustrate the on-sky performance of the Band~9 cartridges. Finally, a
map of the dust and gas distribution around the nearby low-mass protostar
IRAS~16293-2422 \citep{Mundy92, Chandler05, 2011A&A...534A.100J}, taken within
the scope of the science verification, is presented. This dataset is publicly
available.

The on-sky tests also serve to illustrate two of the three specialties that
form the science case of Band~9: very high spatial resolution (the highest of
all four initial bands offered in the Early Science calls for observing
proposals), and high-excitation molecular transitions such as CO $J$=6--5,
which trace warmer gas. The third and last Band~9 science case (not
demonstrated in this work) calls for the capability to observe atomic
fine-structure lines in high-z galaxies, red-shifted into Band~9 wavelengths.


\section{Design and construction}\label{sec:dc}

\subsection{General concept}\label{sec:concept}

\begin{table} 
   \caption{Key specifications for the Band~9 receiver.}
   \centering
   \setlength{\extrarowheight}{3pt}
   \begin{tabular}{ll}
      \hline\hline
      PARAMETER         & SPECIFICATION \\
      \hline
      Frequency range   & 602--720\,GHz \\
      Detector type     & SIS mixer     \\
      Configuration     &
      \begin{minipage}[t]{5.5cm}
         \settowidth{\leftmargini}{\ --\ }
         \begin{itemize}
            \setlength{\itemsep}{0pt}
            \item \raggedright{Simultaneous reception of two orthogonal linear polarizations}
            \item \raggedright{Single-ended DSB operation}
            \item \raggedright{Beam coupling by corrugated feed horn}
            \item \raggedright{Quasi-optical LO insertion by way of beam splitter}
            \item \raggedright{No mechanical tuners}
         \end{itemize}
      \end{minipage} \\
      \noalign{\smallskip}
      Noise temperature &
      \begin{minipage}[t]{5.5cm}
         \settowidth{\leftmargini}{\ --\ }
         \begin{itemize}
            \setlength{\itemsep}{0pt}
            \item \raggedright{Lower than 335\,K (SSB) over 80\% of the band}
            \item \raggedright{Lower than 500\,K (SSB) at any frequency}
         \end{itemize}
      \end{minipage} \\
      \noalign{\smallskip}
      IF coverage       & 4--12\,GHz \\
      \hline
   \end{tabular}
   \label{table:specs}
\end{table}

\begin{figure} 
   \centering
   \includegraphics[width=0.40\textwidth]{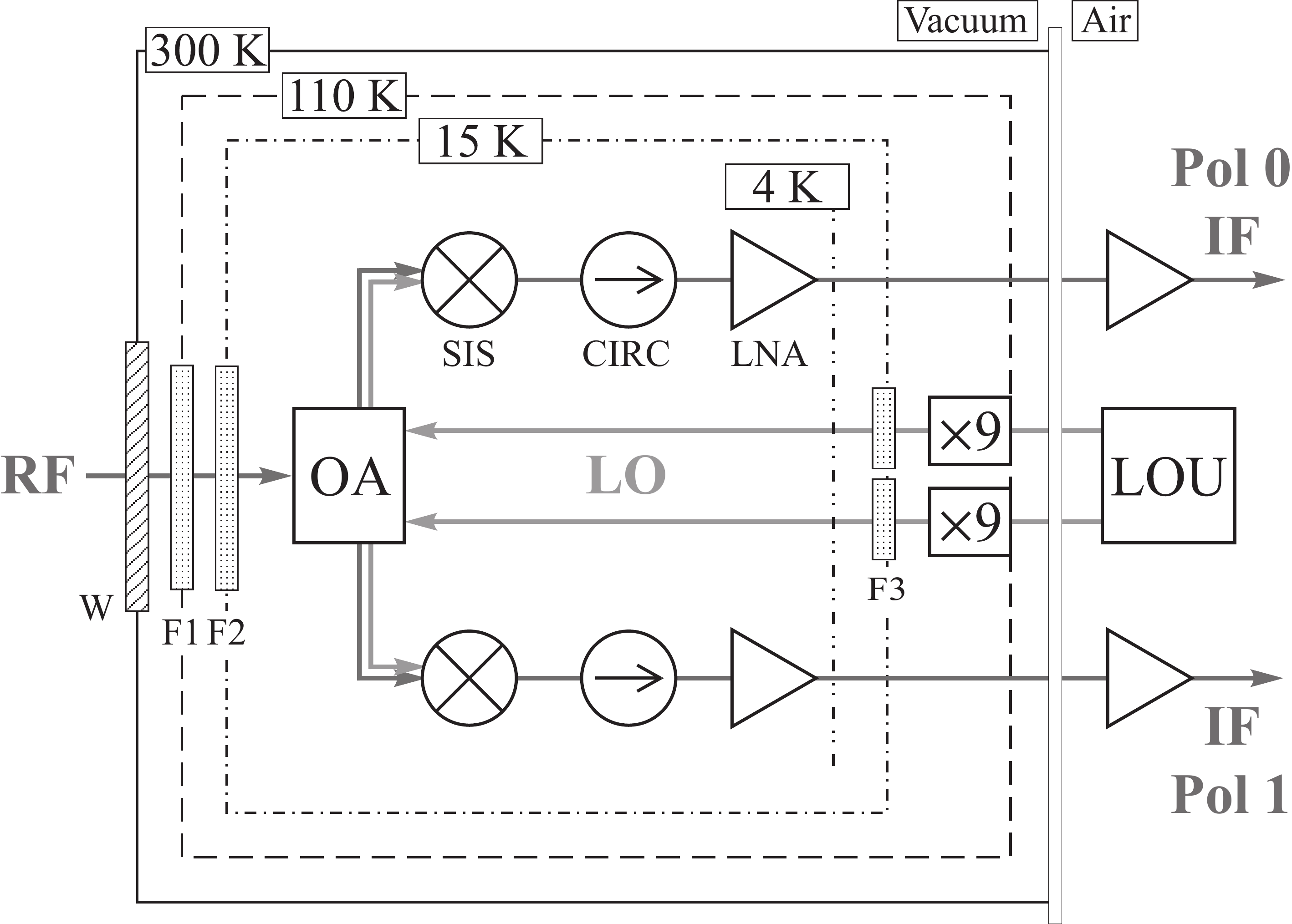}
   \caption{Conceptual design of the Band~9 receiver. The RF signal enters the
            receiver through a cryostat window W and infrared filters F1 and
            F2. In the optics assembly (OA) the beam is refocused as well as
            split into orthogonal polarization components and combined with the
            LO signals. Each polarized beam feeds a SIS-based DSB mixer (SIS),
            and the resulting IF signal passes through a circulator (CIRC)
            before being amplified (LNA) for further processing. The LO signal
            is initially generated at the local oscilator unit (LOU) outside
            the cryostat and then multiplied by nine on the 110\,K stage
            ($\times$9). A pair of infrared blocking filters (F3) is placed in
            front of the LO sources in order to minimize the thermal radiation
            load from the 110\,K level.}
   \label{fig:layout}
\end{figure}

Table~\ref{table:specs} summarizes the key specifications for Band~9, based on
the requirements set by the ALMA Scientific Advisory Committee \citep{wild01,
wild02}. They represent a small selection of the technical goals and guidelines
for the work presented here.

The conceptual design of the Band~9 cartridge is presented in
Fig.~\ref{fig:layout}. The radiation, collected and imaged by the telescope,
enters the cryostat through a vacuum window and a set of infrared filters. The
beam is then refocused in a reflective optics assembly that also contains a
polarizing grid and two beam splitters for, respectively, splitting the light
into two orthogonal linear polarizations and combining each resulting beam with
its LO signal. The two beams are then sent to independent
Superconductor-Insulator-Superconductor (SIS) mixers and subsequently two IF
low-noise amplification chains. The entire receiver is built in a ``cartridge''
that serves as mechanical support and thermal anchor at the different
temperature stages of the cryostat, namely 110, 15 and 4\,K.

In the design it was important to consider reproducibility and ease of
assembly. Particular to ALMA is the necessity of delivering 70+ fully tested
receivers (to fill the main array and provide a couple of spares) in a period
of only four years. In the rest of this section we describe in more detail the
implementation of the components presented in Fig.~\ref{fig:layout}.

\subsection{Optics}

\subsubsection{Window and filters}

Although the window and heat filters are not part of the Band~9 cartridge
proper, they are described here to present an overview of the entire optics
train.

The receiver inside the cryostat looks at the secondary reflector of the
telescope through a vacuum window, which consists of a quartz substrate with an
antireflection coating designed to have an optimized transmission to cover the
frequency range of Band~9 \citep{koller_m397}. In the 110 and 15\,K radiation
shields, infrared blocking filters, consisting of several sheets of expanded
PTFE (Goretex$^{\mbox{\scriptsize\textregistered}}$~\footnote{\raggedright{W.
L. Gore \& Associates, Incorporated, 555 Paper Mill Road, Newark, DE 19711,
USA. Available online:
\url{http://www.gore.com/en_xx/products/sealants/gaskets/gore_gr_sheet_gasketing.html}}}
and Mupor$^{\mbox{\scriptsize\textregistered}}$~\footnote{\raggedright{Interstate
Specialty Products, 55 Gilmore Drive Sutton, MA 01590, USA. Available online:
\url{http://www.interstatesp.com/porex_ptfe_standard_sizes.html}}}), have been
placed \citep{baryshev_m551}.

\subsubsection{RF focusing and LO injection}

\begin{figure} 
   \centering
   \includegraphics[width=0.38\textwidth]{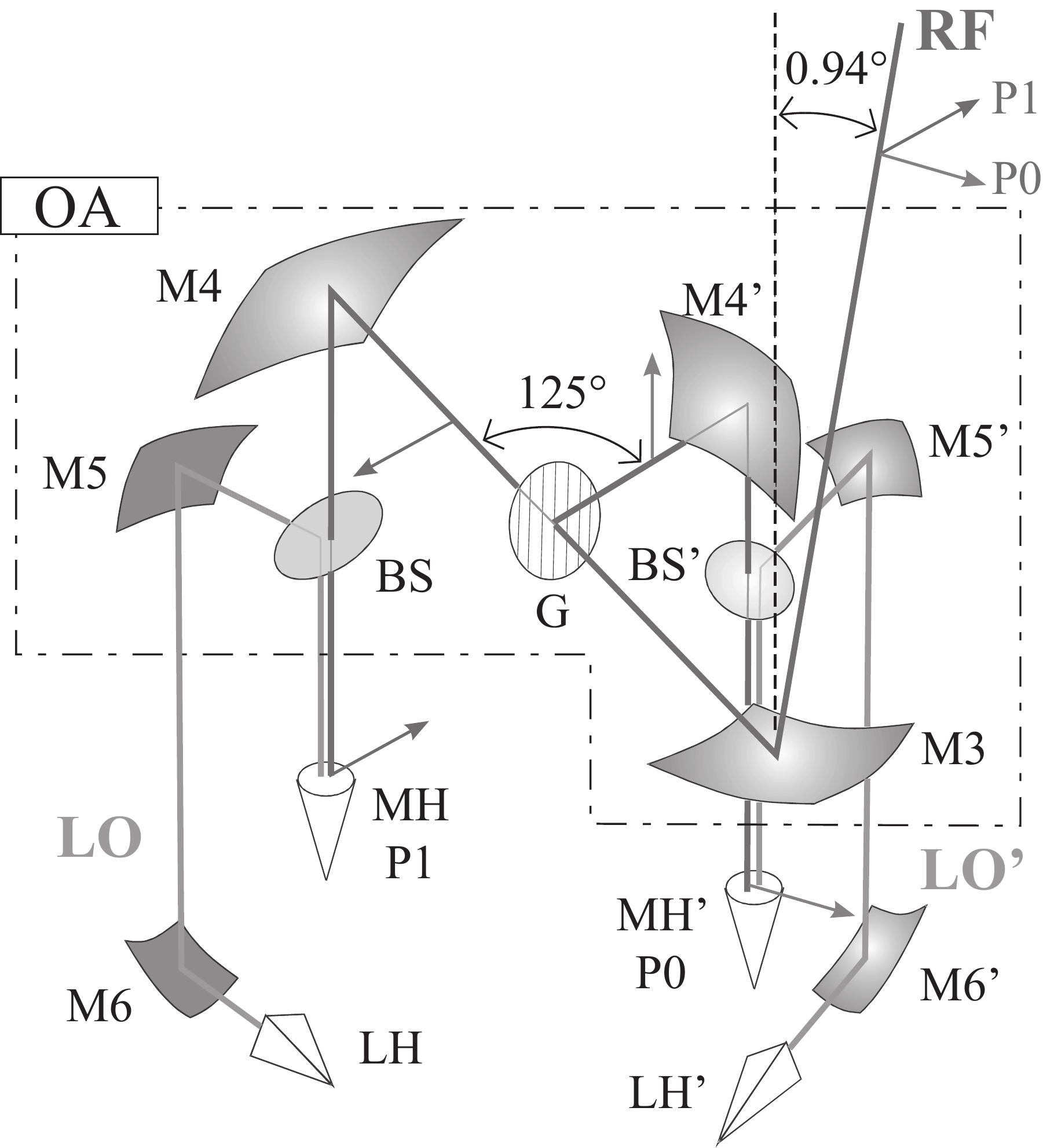}
   \caption{Layout of the optical system for focusing the RF beam and injecting
            the LO signals. The RF beam coming from the telescope hits the
            first in-vacuo mirror, M3. Then, the beam reaches a polarizing
            grid, G, where it is divided in its two linear-polarization
            components, P0 and P1. We consider here the P1 path only; the P0
            path is identical, but with all component labels primed. After the
            polarizer, the beam hits a second mirror M4 that completes the
            focusing of the beam into the mixer horn MH. In front of the mixer
            horn there is a 5\% reflective beam splitter BS, used to insert the
            LO beam that is formed by the LO horn, LH, on the 110\,K level, and
            subsequently focused by the two LO mirrors M6 and M5, into the
            mixer horn. All mirrors are off-axis ellipsoids. Since the
            cartridge is located off the main telescope axis, the incoming RF
            beam has to form an angle of 0.94\degr\ with respect to the
            cartridge axis. The components inside the dashed polygon form the
            optics assembly (OA), situated entirely at the 4\,K level. The
            optics assembly also contains two beam dumps (not shown here) that
            absorb the radiation from the LO source that is transmitted through
            the beam splitters, and one that terminates the unused fourth
            ``port'' of the polarizer.}
   \label{fig:ou-layout}
\end{figure}

The ALMA Band~9 optics is designed to be an integral part of ALMA front end
optics \citep{carter2004optics}. Figure~\ref{fig:ou-layout} shows the layout of
the optical system used to couple the RF and LO signals into the two DSB mixers
corresponding to the two orthogonal linear polarizations. To determine the
focal lengths of the mirrors, a geometrical approximation was used. The sizes
of the mirrors were determined using a Gaussian optics analysis \citep{wild02}.
The sizes were chosen as to fit a 5-waist beam for the RF path, and a 4-waist
beam for the LO. A detailed description of these dimensions is given in
\citet{baryshev_m394}. Furthermore, the whole system was verified using a full
physical optics method \citep{candotti09}.

\begin{figure} 
   \centering
   \includegraphics[width=0.4\textwidth]{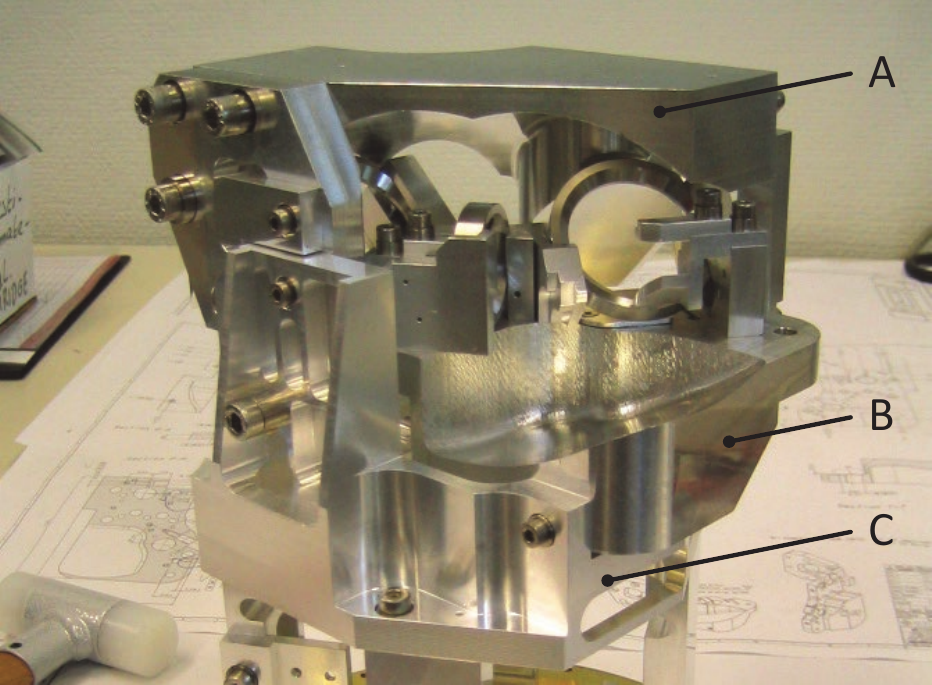} \\
   \vspace{0.4mm}
   \includegraphics[width=0.4\textwidth]{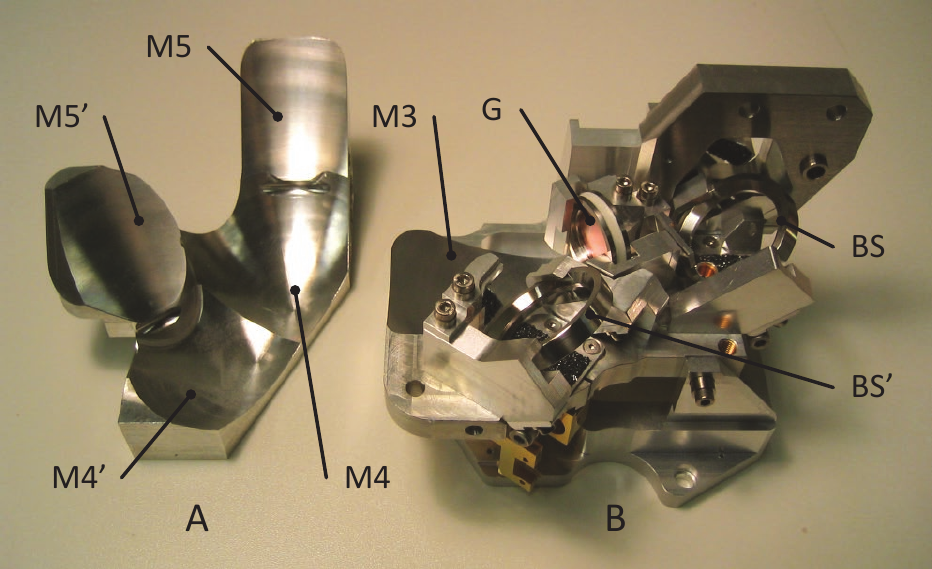}
   \caption{\emph{Top:} implementation of the optics assembly. It consists of
            an upper and a lower block, A and B, mounted in a precision bracket
            or ``cradle'', C. \emph{Bottom:} the upper piece, A, is a single
            aluminum block into which four mirror surfaces (M4, M4', M5 and
            M5') have been machined. The lower piece, B, is also an aluminum
            block where the M3 mirror was machined together with mounting
            structures for the grid (G), beam splitters (BS and BS'), beam
            dumps and mixers. For the sake of clarity, the beam dumps are not
            shown here.}
   \label{fig:ou-implement}
\end{figure}

The 4\,K components of the optical system used to couple the RF and LO signals
are implemented as a single unit, called the optics assembly
(Fig.~\ref{fig:ou-implement}). The optics assembly consists mainly of a pair of
aluminum blocks out of which all of the 4\,K optical surfaces have been
machined. The upper block contains M4, M4', M5 and M5', and is aligned by way
of dowel pins to the lower block, which contains M3. The lower block also
contains aligned attachment points for the grid, beamsplitter and mixer
holders. Alignment accuracy is reached purely by controlled machining
tolerance; there is no provision for adjustment or shimming in the design. This
design demands an accuracy of 40\,$\mu$m in the relative positioning of the
mirror surfaces, which is a value that can be readily achieved with modern
industrial precision machining \citep{baryshev_m394}.

The only two mirrors not on the 4\,K level, M6 and M6', are machined as part of
(identical) LO subassemblies that also contain the mechanical interface to the
LO horn and the final $\times9$ multiplier block, as well as alignment features
matched to the 110\,K plate. The LO horns are of the diagonal pyramidal type.
Although these give a lower efficiency (84\%) than corrugated horns, they are
much easier to manufacture, while the LO produces sufficient power to
compensate for the lower coupling.

The optics assembly has been verified at room temperature using an in-house
built phase-and-amplitude beam pattern measurement setup
\citep{2004ESASP.554..365B} that uses 2D near-field scans. Simulations and
measurements agree extremely well. The main results were \emph{i)} the first
side lobes are at less than $-22$\,dB below the main peak, \emph{ii)} the beam
squint between the polarizations is 9.2\%, and \emph{iii)} the aperture
efficiency is above 80\% at all frequencies and both polarizations
\citep{candotti09}.

\subsection{Mixers}

\subsubsection{Mixer structure}

The mixer (Fig.~\ref{fig:mixer}) consists of a feedhorn, a back piece holding
the SIS mixer device and a spring-loaded ``cap nut'' to clamp the two together.

The mixer horn is of the conical corrugated type, which produces a good
Gaussian beam with a high coupling efficiency (98\%). To facilitate its
production, it was designed to have the shortest slant length possible without
degrading its beam pattern, which reduces the number of necessary corrugations
\citep{baryshev_m394}.

The back piece contains the SIS device with backshort cavity, a miniature
coaxial GPO connector to export the IF signal, and a small heater resistor (the
``deflux heater''), enabling the mixer temperature to be lifted above $T_c$
temporarily in order to get rid of magnetic flux trapped in the superconducting
film near the SIS junction.

The back piece is aligned with the horn by an accurately machined centering
ring. Two pole pieces penetrate the ring to bring the magnetic field as close
as possible to the junction. The pole pieces also take care of the angular
alignment of the back piece to the horn.

\subsubsection{Mixer mounting}

\begin{figure} 
   \centering
   \includegraphics[width=0.4\textwidth]{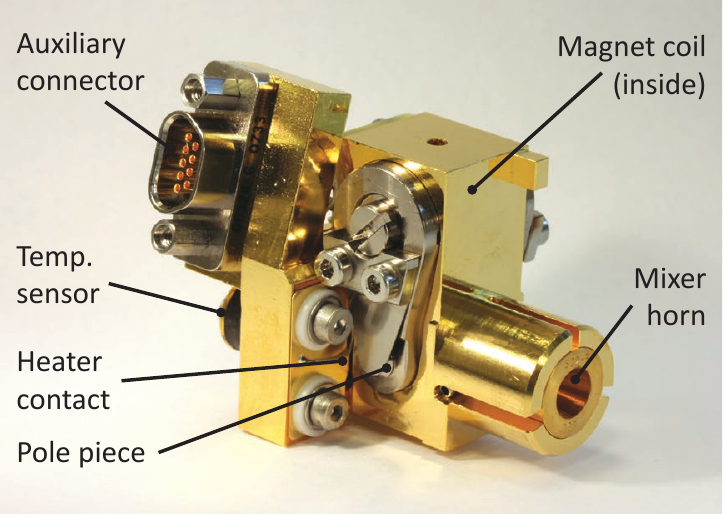} \\
   \vspace{0.4mm}
   \includegraphics[width=0.4\textwidth]{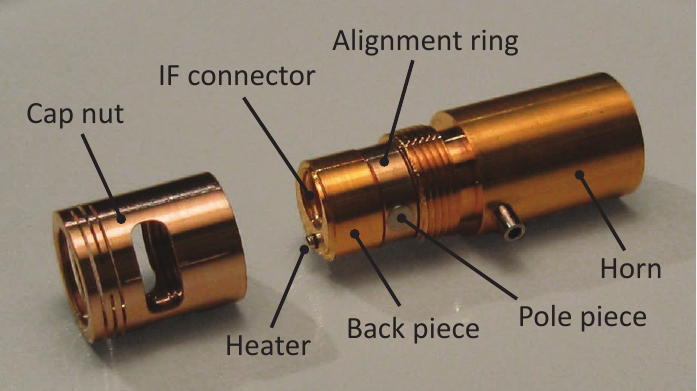} \\
   \vspace{0.4mm}
   \includegraphics[width=0.4\textwidth]{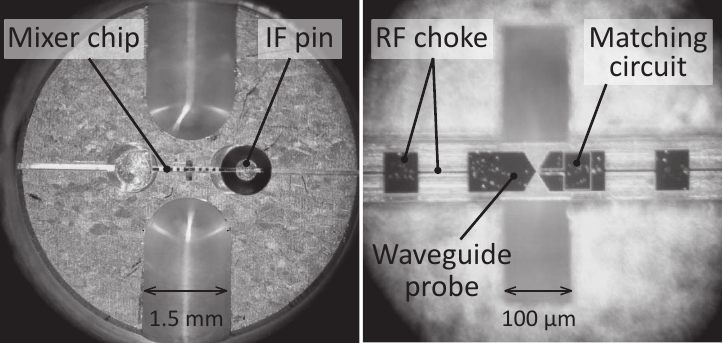}
   \caption{\emph{Top:} mixer in mixer holder. \emph{Middle:} detailed view of
            the mixer with the ``cap nut'' removed. \emph{Bottom:} detailed top
            view of the back piece containing a mixer chip (left) and close-up
            of the central part of the mounted mixer chip (right).}
   \label{fig:mixer}
\end{figure}

The mixer, as described above, is mounted into a mixer holder
(Fig.~\ref{fig:mixer}) that, in turn, is clamped into the optics assembly. The
mixer holder contains a superconducting magnet coil for Josephson-current
suppression and magnetic flux conductors to bring the field to the inner pole
pieces of the mixer. When inserting the mixer into the holder, the GPO IF
connector in the mixer back piece is blind-mated to a GPO-to-SMA adapter by way
of a coaxial ``bullet'', and at the same time, the deflux heater in the back
piece is contacted by a spring contact. Finally, the mixer holder contains a
Si-diode temperature sensor.

\subsection{Superconducting tunnel devices}

For Band~9 (602--720\,GHz) the decision was taken to use an all-niobium
technology, rather than the more complex technology used for Herschel's HIFI
Band~3 and Band~4, in which niobium tunnel junctions are integrated with a
highly conductive aluminum microstrip over a groundplane of the high-gap
superconductor NbTiN \citep{jackson2001}. Since the superconducting energy gap
of niobium of (at best) 2.9\,meV corresponds to 700\,GHz, no deterioration of
niobium by poor deposition conditions could be tolerated (for instance, the
$T_c$ and superconducting gap of niobium depend strongly on built-in oxygen).

The mixer chip consists of a 40\,$\mu$m quartz substrate on which the SIS
tunnel junctions and impedance networks are processed. The superconducting
circuitry consists of a symmetrical bow-tie waveguide probe to couple the
radiation coming from the horn, a stripline impedance-matching network between
the probe and the SIS junction, followed by a path to transport the IF signal
and DC bias. This IF signal path also contains an RF choke structure to prevent
the RF signal to leak out to the IF channel. The mixer chip is located in front
of a fixed waveguide backshort to maximize the electromagnetic coupling. The
dimensions of the probe and the backshort were optimized using Ansoft
HFSS\footnote{Ansoft Corporation, Four Station Square, Suite 200, Pittsburgh,
PA~15219-1119, USA.} and CST Microwave Studio\footnote{Microwave Studio,
Connecticut, MA, USA (2006). Available online: http://www.cst.com/.}. The same
was done, though independently, to optimize the dimensions of the RF choke
structure and the channel where the chip is mounted.

The coverage of the full band is first of all determined by the RC time
constant of the tunnel junction. Since both R and C scale with the area, the RC
bandwidth can not be increased by changing the lateral dimensions, but only by
using a tunnel barrier as thin as possible, taking advantage of the fact that
the resistance depends exponentially on thickness (it is a quantum-mechanical
tunneling process), while the capacitance depends linearly. This requirement
leads to tunnel barriers with thicknesses in the 1--2\,nm range, with typical
resistance per area values for aluminum-oxide tunnel barriers of maximally
25\,$\Omega\mu$m$^2$. The area of the SIS junctions is chosen to provide a
termination in a $\approx\!\!2R_N$ load. Since the chips will be connected to a
50\,$\Omega$ output, the aim in the fabrication of the SIS junctions was to
have a normal resistance of 25\,$\Omega$. Fixing the resistance of the junction
and its area and knowing the input impedance of the waveguide probe, the
dimensions of the matching network can be calculated. Standard transmission
line theory was employed, using the theory of the anomalous skin effect for the
complex conductivity of Nb \citep{lodewijk07}. This technology was used for the
first eight cartridges in 2009.

Using the existing aluminum oxide tunnel barrier technology as the baseline for
the coverage of Band~9, with the lowest possible noise temperature, remained a
challenge. The upper part of the band could potentially be improved by the use
of Herschel's HIFI technology, using NbTiN for the impedance network. A more
pressing problem, however, was that a better coverage of the atmospheric band
would require even thinner tunnel barriers, which is not reliably achievable
with aluminum oxide and therefore would cause a rapid decline in production
yield. In view of some indications that aluminum nitride barriers might allow
lower $R_nA$ values without deterioration of the quality of the non-linear
current-voltage characteristics, a program was started to develop such a tunnel
barrier process \citep{zijlstra07} using a plasma source to create the
chemically active nitrogen radicals. The success of this development led to the
inclusion of this device technology in the remaining Band~9 cartridges.

\subsubsection{SIS device fabrication}

The AlO$_x$ barrier SIS devices are fabricated on a 200\,$\mu$m thick quartz
substrate \citep{lodewijk07, mena-mtt, zijlstra07}. First, a sacrificial Nb
monitor layer is deposited, followed by an optically defined trilayer of
Nb/Al/AlO$_x$/Nb. The thickness of the Al layer is 5--7\,nm, the top and bottom
niobium layers are 100\,nm thick. Junctions are defined by electron-beam
lithography using negative SAL601 resist and etched in a mixture of
SF$_6$/O$_2$ reactive ion etch (RIE) plasma. The AlO$_x$ layer acts as an etch
stop for the ground plane. The junction resist pattern is subsequently used as
a lift-off mask for a sputter-deposited dielectric layer of SiO$_2$. A Nb/Au
top layer is deposited to define the top wire and contact pads. The Au is
wet-etched in a KI/I$_2$ solution using an optically defined window. Finally,
using an e-beam defined top wire mask pattern, the top layer of Nb is etched in
a SF$_6$/O$_2$ RIE plasma, which finishes the fabrication process.

An identical process is used for junctions with AlN tunnel barriers, except for
the barrier growth itself. AlN is formed by exposing the Al over-layer to the
after-glow region of a nitrogen plasma from an inductively coupled plasma
source. The source is operated at a pressure of $4\times10^{-2}$\,mbar and a
power set point of 550\,W; source to wafer distance is fixed at 10\,cm. Using
nitridation time as a control parameter, we made Nb/Al/AlN/Nb trilayers
yielding critical current densities in the range of 20--40\,kAcm$^{-2}$.

\subsubsection{SIS performance verification}

The fabrication process results in hundreds of devices on a single wafer. After
dicing the wafer, the mixer chips are first tested in a dip stick for their I-V
characteristics. Those showing good results are mounted in a DSB mixer block
which is installed inside a small test cryostat, where they are tested for
frequency band coverage using a Fourier transform spectrometer. When the band
coverage is found to be adequate, the devices are transferred to a heterodyne
test cryostat. This cryostat uses the same window and IF chain as the cartridge
cryostat, but the LO is generated at room temperature. The junctions showing
the best noise temperatures are chosen to be installed in the receiver
cartridges. On average, the measured noise temperatures in the test cryostat
were about 20\,K higher than those obtained when they are integrated in the
receiver and tested in the cartridge cryostat. The improved noise temperature
in the cartridge is mostly attributable to the reduced contribution of the
black-body background radiation from the final LO multiplier, due to its lower
physical temperature (110\,K vs. 300\,K) and the thinner beam splitter
(6\,$\mu$m vs. 12\,$\mu$m) employed in the cartridge. In both cases, the
electronic noise of the LO itself is expected to be comparable.

\subsection{IF chain}

The IF output of the mixer first passes through an isolator and then to a low
noise amplifier (LNA), both on the 4\,K stage of the cartridge. The 4--12\,GHz
isolator (circulator) was developed under contract by Pamtech\footnote{Passive
Microwave Technology, Inc., Camarillo, CA, USA. [Now part of QuinStar
Technology, Inc., Torrance, CA, USA]} and the amplifier at Centro Astron\'omico
Yebes \citep{lopez06}. The amplifier has three stages of amplification using
InP high electron-mobility transistors. Finally, an extra amplification of the
IF signal is done immediately outside the cold cartridge using commercial
amplifiers, prior to being exported to the telescope's IF system. The IF cables
between the temperature stages are made out of stainless steel and are
thermally anchored at each stage to minimize heat input to the 4\,K system. On
the 12\,K stage there are 3--5\,dB attenuators which are matched to the
respective LNA gains in order to equalize the IF output powers between
cartridges.

\begin{figure}[t] 
   \centering
   \includegraphics[width=0.48\textwidth]{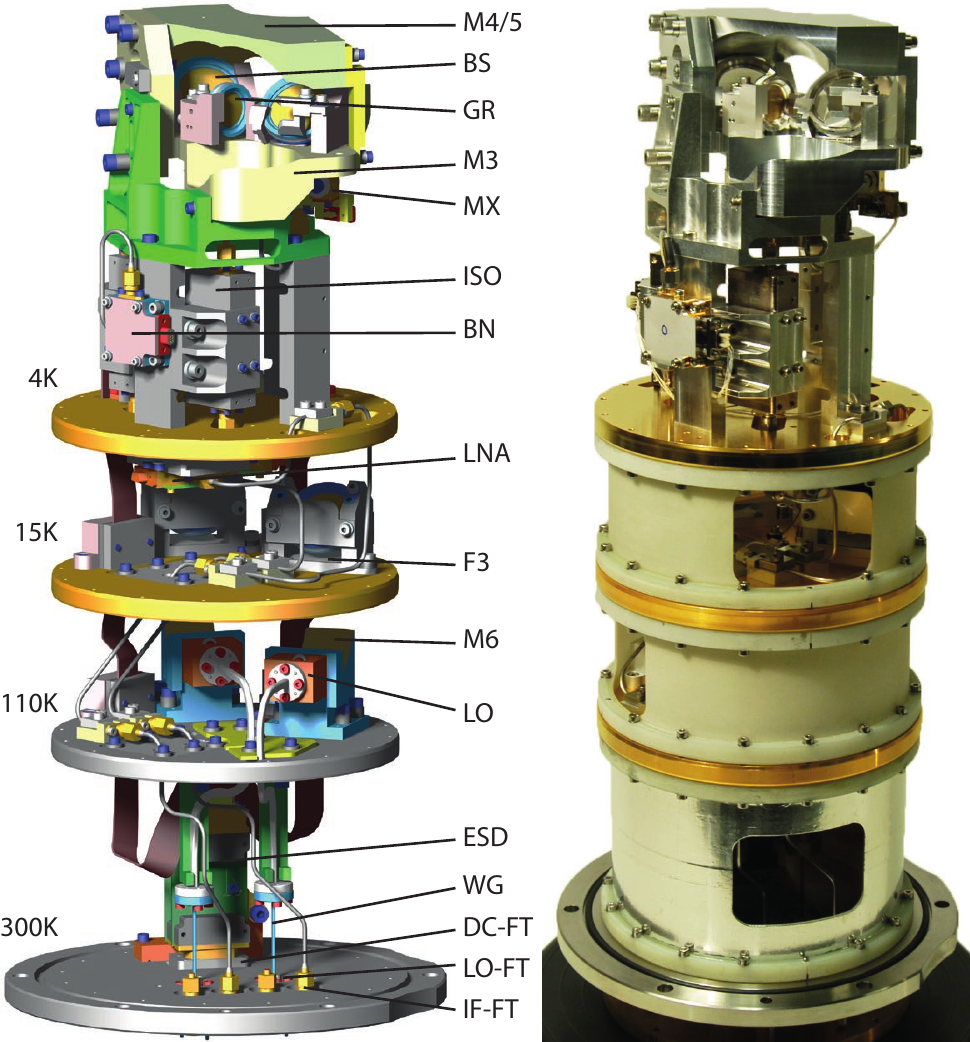}
   \caption{CAD drawing (with the fiberglass spacer rings omitted) and
            photograph of the assembled receiver. The optics assembly contains
            the top part (M4/5), bottom part (M3), beam splitter (BS),
            polarizing grid (GR) and the mixers (MX). Further down on the 4\,K
            level are the bias network boxes (BN), isolators (ISO), and LNAs
            (LNA). On the 15\,K level are the LO infrared blocking filters
            (F3). The 110\,K level contains the LO multipliers (LO) with
            associated mirrors (M6). Finally, on the 300\,K level (the vacuum
            flange) reside the ESD protection board (ESD) on top of the DC
            feedthroughs (DC-FT), and the LO feedthroughs (LO-FT) and IF
            feedthroughs (IF-FT). The LO waveguide (WG) contains a
            stainless-steel thermal break.}
   \label{fig:integra}
\end{figure}

\subsection{LO unit}

The LO signal is generated by a YIG oscillator and an electronically tunable
solid-state chain of power amplifiers and multipliers \citep{bryerton05}. The
YIG and the first stage of the chain, a multiplier ($\times3$) and a power
amplifier, are placed outside the cryostat in the LO unit. This unit also
contains a PLL system that locks the YIG signal to the photonic reference
signal distributed to the ALMA antennas.

Since it was considered impractical to couple a 600 to 720\,GHz signal from the
air side into the vacuum space (either quasi-optically or through a waveguide),
and also because of noise considerations, the final $\times9$ multiplication is
done in vacuum, on the 110\,K stage \citep{hesper05}. A WR12 waveguide
transports the 66--80\,GHz signal to the multiplier block, using a thin mica
window as a vacuum barrier.

\subsection{Integration}

The different components described in the previous sections are integrated in a
standard ALMA cartridge body according to the layout presented in
Sect.~\ref{sec:concept}. Figure~\ref{fig:integra} presents the integrated
receiver cartridge. The assembly is performed from bottom to top and the
cartridge construction allows for easy access to most of the critical
components in case of service or repair. All components can be built in and out
without need of mechanical adjustment, since alignment is achieved by machining
tolerances only. In particuler, the mixers can be replaced in very short time
without any major disassembly.


\section{Laboratory results}\label{sec:lab}

\subsection{Performance}

For laboratory testing, every fabricated receiver is introduced in a
single-cartridge ALMA test cryostat \citep{sekimoto_m455}. An automated
measurement system \citep{barkhof2009alma}, based on scripted measurement
routines, has been built to perform the receiver characterization in a uniform
and consistent way for 73 Band~9 cartridges.

Several important receiver characteristics such as sensitivity, gain
compression, receiver stability and beam parameters have been characterized and
results are presented in following sections. Where statistical values are
presented, these apply to 15 receivers containing AlO$_x$ junctions and up to
52 receivers containing AlN junctions.

\subsubsection{Receiver beam parameters}

\begin{figure}[t] 
   \centering
   \includegraphics[width=0.46\textwidth]{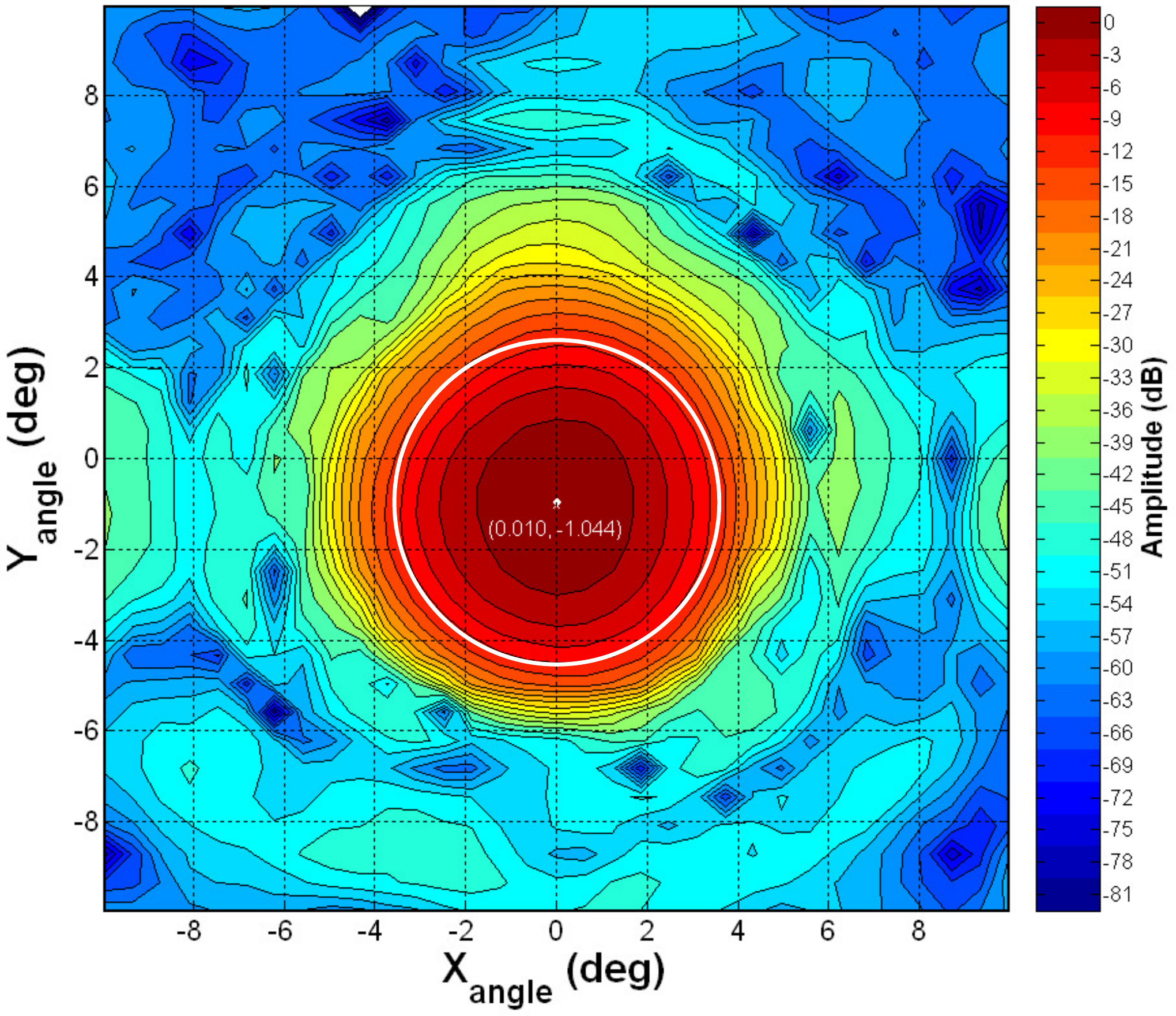}
   \caption{Typical co-polar beam pattern in a Band~9 cartridge, with the
            angular extent of the secondary telescope mirror indicated (white
            circle). This pattern corresponds to polarization~0 in cartridge
            \#63.}
   \label{fig:beam}
\end{figure}

The receiver beam parameters are determined by measuring 2D near-field phase
and amplitude scans for both polarization channels. The measurement system is
conceptually similar to the one described in \citet{carter2002phase,
baryshev2004verification} and consists of a 3D scanner, which is mounted above
the test cryostat; a beam probe, utilizing a corrugated feedhorn integrated
with a harmonic generator based on a superlattice device
\citep{paveliev2005frequency} and microwave circuitry to create a reference
signal. Phase and amplitude in each scan point are registered by a vector
network analyzer. 2D beam maps are scanned in two planes perpendicular to the
beam propagation axis and offset along the propagation axis by a quarter
wavelength of the measurement frequency. The final beam map is constructed out
of these two scans by adding them with a 90 degree phase shift, allowing
numerical compensation of the parasitic signals associated with standing waves
between the probe and SIS mixer. The far field beam map is then obtained from
the near field data by a complex Fourier transform. An example of such a far
field beam map is presented in Fig.~\ref{fig:beam} with the angular extent of
the secondary mirror shown as a white circle. Most of the beam parameters such
as aperture efficiency, beam squint beam propagation angles can be readily
determined from this far field map \citep{baars_m456}. The typical dynamic
range in the data is 70\,dB, which allows for high accuracy data reduction. The
statistical distribution of measured beam parameters for a large set of ALMA
Band~9 receivers is shown in Table~\ref{table-beam}.

\begin{table} 
   \caption{Average values and standard deviations of beam parameters over 62
            Cartridges.}
   \centering
   \setlength{\extrarowheight}{1pt}
   \begin{tabular}{llr@{\ }l}
      \hline\hline\noalign{\smallskip}
      PARAMETER                            & SYMBOL            &\multicolumn{2}{c}{VALUE} \\
      \hline\noalign{\smallskip}
      \multirow{2}{*}{Angle (\degr)}       & $\theta_{xFF}$    & $ 0.093$ &$\pm$ 0.075 \\
                                           & $\theta_{yFF}$    & $-0.093$ &$\pm$ 0.075 \\
      \hline\noalign{\smallskip}
      \multirow{4}{*}{Focus Position (mm)} & $x$               & $-0.874$ &$\pm$ 0.504 \\
                                           & $y$               & $-1.488$ &$\pm$ 0.271 \\
                                           & $z$               & $76.766$ &$\pm$ 3.004 \\
                                           & $\Delta$          & $ 0.140$ &$\pm$ 0.075 \\
      \hline\noalign{\smallskip}
      \multirow{6}{*}{Efficiencies (\%)}   & $\eta_{spill}$    &  $95.47$ &$\pm$ 0.81  \\
                                           & $\eta_{taper}$    &  $87.82$ &$\pm$ 1.47  \\
                                           & $\eta_{phase}$    &  $99.79$ &$\pm$ 0.10  \\
                                           & $\eta_{polar}$    &  $97.97$ &$\pm$ 0.23  \\
                                           & $\eta_{focus}$    &  $99.91$ &$\pm$ 0.10  \\
                                           & $\eta_{aperture}$ &  $81.89$ &$\pm$ 0.90  \\
      \hline\noalign{\smallskip}
      Beam Squint (\%)                     & ---               &  $ 4.34$ &$\pm$ 2.29  \\
      \hline
   \end{tabular}
   \label{table-beam}
\end{table}

The cross polarization characteristics of the receivers are measured by
inserting a motorized polarizing grid into the receiver beam which, depending
on wire orientation, couples either a 300\,K or 77\,K calibration load signal
into each polarization channel. By rotating the polarizer by 360 degrees and
measuring the output signal of the receiver at the same time one obtains the
cross polarization value for the receiver. A typical cross polarization
response has been found to be $-17$ to $-18$\,dB for all cartridges.

\subsubsection{Noise temperature}

\begin{figure} 
   \centering
   \includegraphics[width=0.44\textwidth]{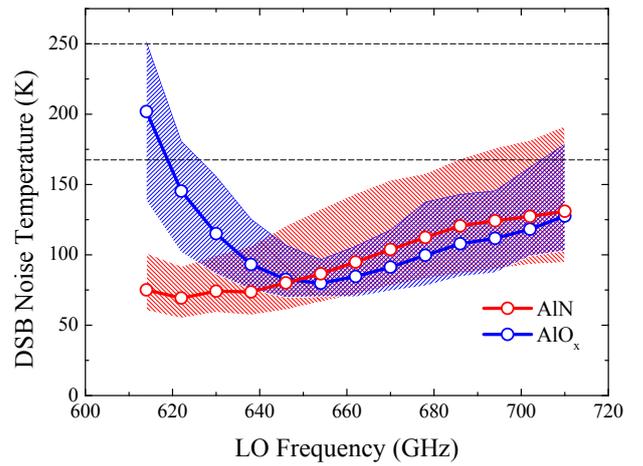}
   \caption{Average of the best noise temperature at different LO frequencies.
            The average corresponds to 15 receivers containing AlO$_x$
            junctions (blue) and 42 receivers containing AlN junctions (red).
            The shaded areas around the average values represents the range
            where the noise temperatures are located. The dashed horizontal
            lines correspond to ALMA specifications according to
            Table~\ref{table:specs}.}
   \label{fig:tnoiseLO}
\end{figure}

The noise temperatures are measured using the standard Y-factor method, with
hot and cold loads at room temperature and 77\,K, respectively.
Figure~\ref{fig:tnoiseLO} presents the best noise temperature averaged over a
large ensamble of receivers at 12 different LO frequencies. The noise in this
figure is integrated over the full 4--12\,GHz IF band. It is apparent from the
figure that AlN junctions cover the band more effectively as they have higher
critical current densities than AlO$_x$ junctions while still having DC I-V
characteristics of sufficient quality \citep{zijlstra07}. In
Fig.~\ref{fig:tnoiseLO} we also present the statistical spread for all
cartridges, demonstrating that all the receivers perform better than the ALMA
specifications (cf. Table~\ref{table:specs}) with margin.

\begin{figure} 
   \centering
   \includegraphics[width=0.44\textwidth]{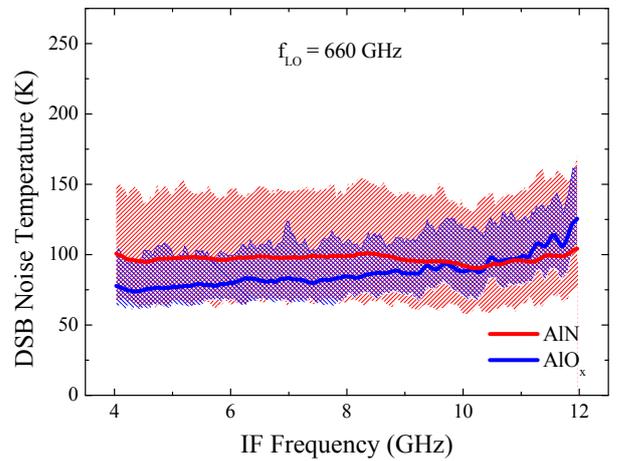}
   \caption{Average of the noise temperature as a function of IF for an LO
            frequency of 660\,GHz. The average corresponds to the 15 receivers
            containing AlO$_x$ junctions (blue) and 52 receivers containing AlN
            junctions (red). The shaded areas around the average values
            represent the range where the noise temperatures are located.}
   \label{fig:tnoiseIF}
\end{figure}

To determine the noise temperature as a function of IF, an electronically
tunable YIG filter is used at the end of the IF chain. The measured IF noise is
shown in Fig.~\ref{fig:tnoiseIF}. Covering such an extremely large frequency
range presents a significant technical challenge. As seen in the figure,
AlN-barrier based SIS mixers perform better than AlO$_x$-barrier junctions at
higher IF frequencies, which can be explained by the smaller parasitic parallel
capacitance and lower output impedance achieved for AlN tunnel barrier
junctions.

\subsubsection{Gain compression}

\begin{figure} 
   \centering
   \includegraphics[width=0.44\textwidth]{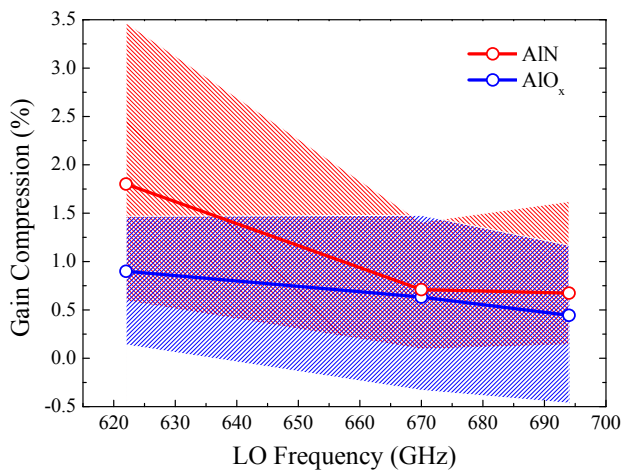}
   \caption{Average gain compression at three LO frequency values. The shaded
            area represents the range of measured gain compression values.}
   \label{fig:gainc}
\end{figure}

We have measured the gain compression using the method described in detail in
\citet{baryshev2004verification}. The receiver gain was measured by inserting a
small, varying, chopped, signal into the receiver beam on top of the large
background created by means of a motorized polarizing grid reflecting either
77\,K or 373\,K black-body radiation into the receiver beam. The small varying
signal is than synchronously detected by a lock-in amplifier allowing high
signal-to-noise measurements. The average gain compression over 67 receivers
and its statistical variation is presented in Fig.~\ref{fig:gainc}. Gain
compression does not depend strongly on the type of junction since it is mostly
determined by the embedding IF impedance seen by the SIS junction and by the
operation frequency \citep{kerr_m401}. All measured values lie within the 4\%
gain compression range for 373\,K load temperature as required by ALMA
specifications.

\section{First-light and scientific results}\label{sec:science}

\subsection{Commissioning}

During the commissioning of ALMA from the fall of 2010 to the spring of 2012, a
large number of tests were performed to characterize the performance of Band~9
at the Array Operations Site (AOS), both on sky and on loads. In this section,
several of the tests performed on sky are presented to demonstrate the
successful commissioning and current performance characteristics of the Band~9
receivers. All tests were performed using the, at that time, most up-to-date
available array at the AOS, consisting mostly of Vertex antennas with,
occasionally, a few Melco and AEM antennas integrated into the array.

First a quasar was tracked for one hour to show the phase stability. The quasar
in question is J1924-292. Then a test in which four quasars around the bright
quasar J0538-441 were observed, is presented. The four quasars are J0538-440,
J0423-013, J0730-116 and J0522-364. We also show some of the results derived
from the Science Verification data sets on both the calibration targets and the
science target, IRAS~16293-2422, taken during commissioning time to verify
Band~9 performance. For all tests, the number of antennas shown is limited, but
the performance on most other antennas are very similar.

\subsection{Quasar tracking}

\begin{figure}[t] 
   \centering
   \includegraphics[width=0.46\textwidth]{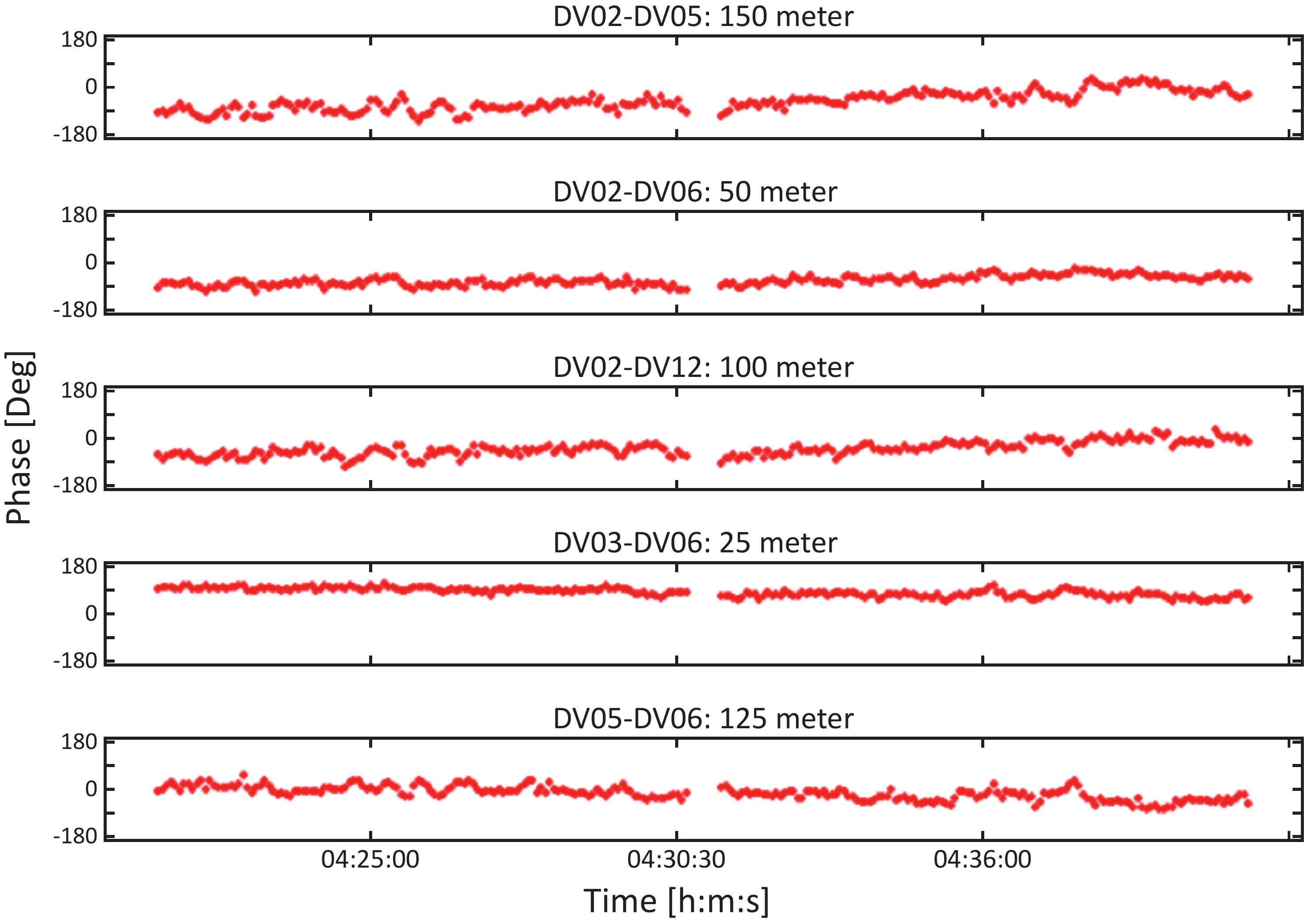} \\
   \vspace{2mm}
   \includegraphics[width=0.46\textwidth]{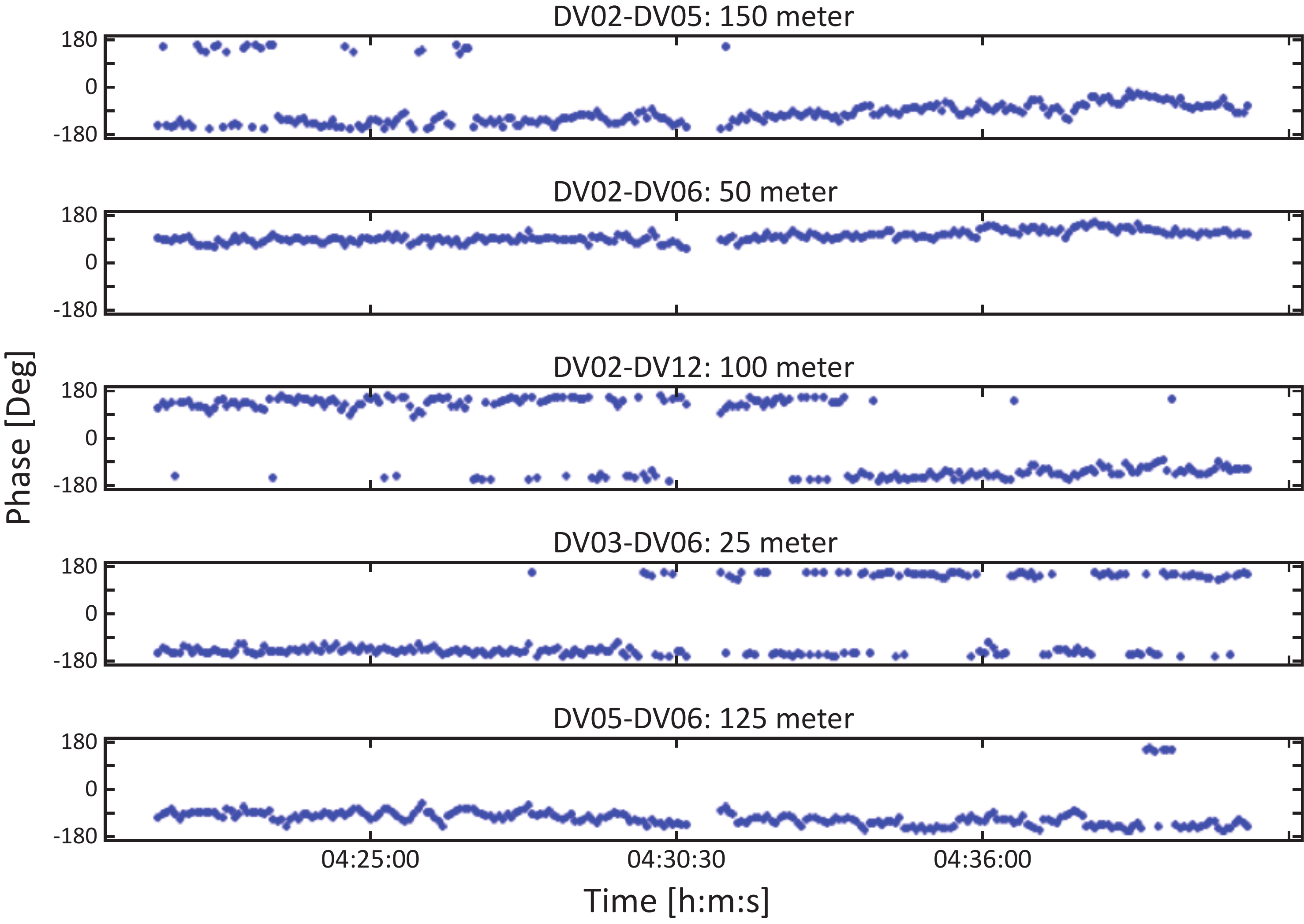}
   \caption{Phase stability of quasar J1924-292 in XX polarization (top five
            panels, red) and YY polarization (bottom five panels, blue),
            tracked over $\approx$20\,min at different selected baselines.
            Physical baseline length is indicated above each plot.}
   \label{fig:phasexx1924}
\end{figure}

\begin{figure}[t] 
   \centering
   \hbox{\hspace{2.2mm}\includegraphics[width=0.482\textwidth]{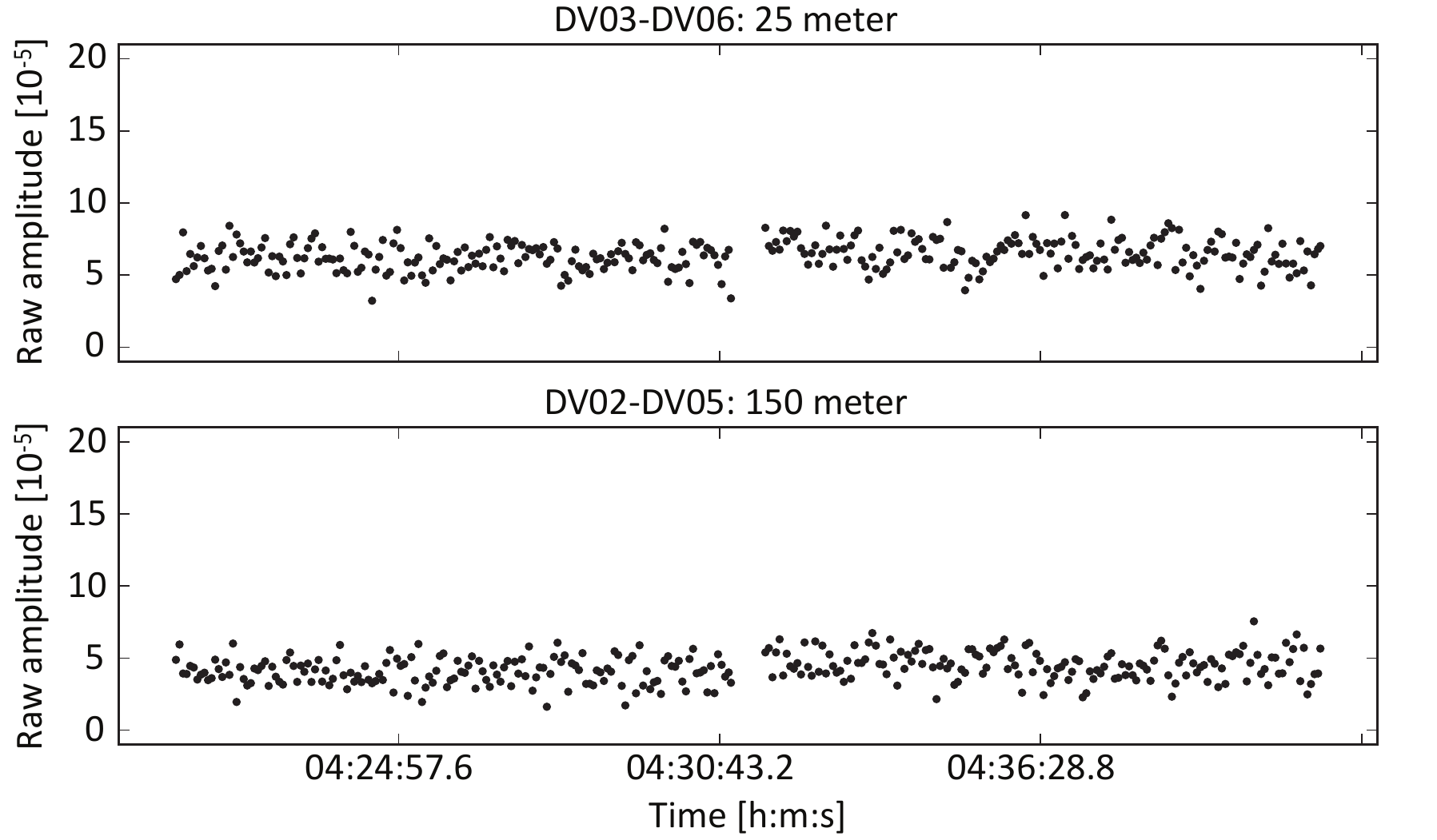}}
   \caption{Amplitude stability of quasar J1924-292 on the shortest and longest
            tested baselines.}
   \label{fig:amp1924}
\end{figure}

\newcommand{\cc}{\multicolumn{1}{c}}
\newcolumntype{d}{D{.}{.}{-1}}
\begin{table*}[t] 
   \caption{Phase and amplitude stability results on J1924-292 taken on
            September 11, 2011.}
   \centering
   \setlength{\extrarowheight}{1pt}
   \begin{tabular}{cdddddd}
      \hline\hline\noalign{\smallskip}
      Baseline  &\cc{Physical length} &\cc{Projected Length} &\cc{Mean Phase} &\cc{Phase RMS} &\cc{Mean Amplitude} &\cc{Amplitude RMS}   \\
                &\cc{m}               &\cc{m}                &\cc{deg}        &\cc{deg}       &\cc{\ \ \ 10$^{-5}$}&\cc{\ \ \ 10$^{-5}$} \\
      \hline\noalign{\smallskip}
      DV03-DV06 & 35.8                & 24.2                 & 79.2           &42.4           & 8.67               & 1.50                \\
      DV02-DV06 & 63.8                & 58.7                 &-88.0           &35.4           & 7.23               & 1.50                \\
      DV05-DV06 &113.9                & 69.5                 &-16.8           &26.6           & 7.15               & 1.41                \\
      DV02-DV12 &123.4                &101.7                 &-69.3           &27.2           & 5.56               & 1.54                \\
      DV02-DV05 &159.4                &110.5                 &-74.8           &32.1           & 5.79               & 1.51                \\
      \hline
   \end{tabular}
   \label{table:mean}
\end{table*}

On September 11, 2011, ALMA observed the quasar J1924-292 (RA=19:24:51.056,
Dec=$-29$:14:30.121, J2000) for 20 minutes in Band~9 using 13 antennas in dual
polarization to track the phase and amplitude stability. At the time of
observing, J1924-292 was one of the brightest quasars in the sky at these high
frequencies. Although no actual flux measurement was done at 690\,GHz at this
time, comparison from previous measurements in Band~9, and extrapolations from
Band~3 and~6 measurements a week before and after, an estimate of the flux at
around 4\,Jy with an error of 0.5\,Jy could be made. The phases and amplitudes
on a large number of different baselines are inspected. At the time of
observations, commissioning of several modes was still ongoing. The analysis
has been limited to a few selected baselines. The baselines are chosen to
sample projected baseline lengths of the array, which was being built up toward
the start of Early Science and thus consist mostly of short ($<$200\,m)
baseline lengths. Five baselines were selected, ranging from 25 to 150\,m, with
projected baselines of 24.2 to 110.5\,m. The results can be seen in
Fig.~\ref{fig:phasexx1924}, which shows the phase stability in the two
polarizations. The precipitable water vapor during the observation was measured
at 0.2$\pm$0.02\,mm and very stable conditions. The target was between 35 and
40 degrees elevation, relatively far from the highest transit. All phases are
extremely flat with RMS numbers of $\sim$30--35 degrees. Only the short
baseline was slightly higher due to a very small delay trend in antenna DV03.
The amplitudes also agree very well with each other and the RMS in the
uncalibrated amplitudes is almost equal at all baseline lengths
(Fig.~\ref{fig:amp1924}).

The mean phase and amplitude are summarized in Table~\ref{table:mean}. All
numbers are given uncalibrated and uncorrected from the radiometer. Corrections
using the water vapor radiometer solutions, derived using the WVRGCAL software
\citep{Nikolic12}, were applied in the test and found to only marginally
improve the data, likely due to the very dry conditions. The overall stability
is very good (phase noise smaller than $\pm45\degr$, amplitude noise
$1.5\times10^{-5}$ normalized to the peak flux of the quasar) over the half
hour test with excellent stability in all receivers. When taking the brightness
of the source, baseline lengths and weather into consideration, the spread in
the phase and amplitude RMS is dominated by the remaining atmospheric
conditions. In addition, since five unique cartridges were used in the
analysis, the stability of the receivers compared to each other is very
similar. Comparison was done to the cartridges not displayed and the same
results were obtained for all cartridges.

\subsection{Four-quasar experiment}

In the southern sky, a set of four relatively bright quasars can be found
within an hour angle of approximately three hours, at similar declinations.
These four quasars are J0538-440 (3.0\,Jy, Oct 13, 2011), J0423-013 (2.2\,Jy
Oct 1, 2011), J0730-116 (0.5\,Jy, Oct 13, 2011) and J0522-364 (4.1\,Jy, Sep 9,
2011). Fluxes are given at 343\,GHz (Band~7) with the date of measurement
closest to the measurement taken for phase transfer. Although these quasars are
significantly weaker at short wavelengths, they are all detectable in Band~9.
An experiment was done on October 6th, 2011 to observe this set for two hours.
Each quasar was observed for 60 seconds before looping to the next. This entire
sequence was then repeated ten times. This test is aimed to demonstrate phase
transfer between sources across the sky, as well as the ability to apply phase
solutions of the brightest source (in this case J0522-364 or J0538-440) to the
other sources. The configuration, and thus the distribution of baselines, were
similar to those in the stability experiment described above.

\begin{figure} 
   \centering
   \hspace{-4mm}\includegraphics[width=0.44\textwidth]{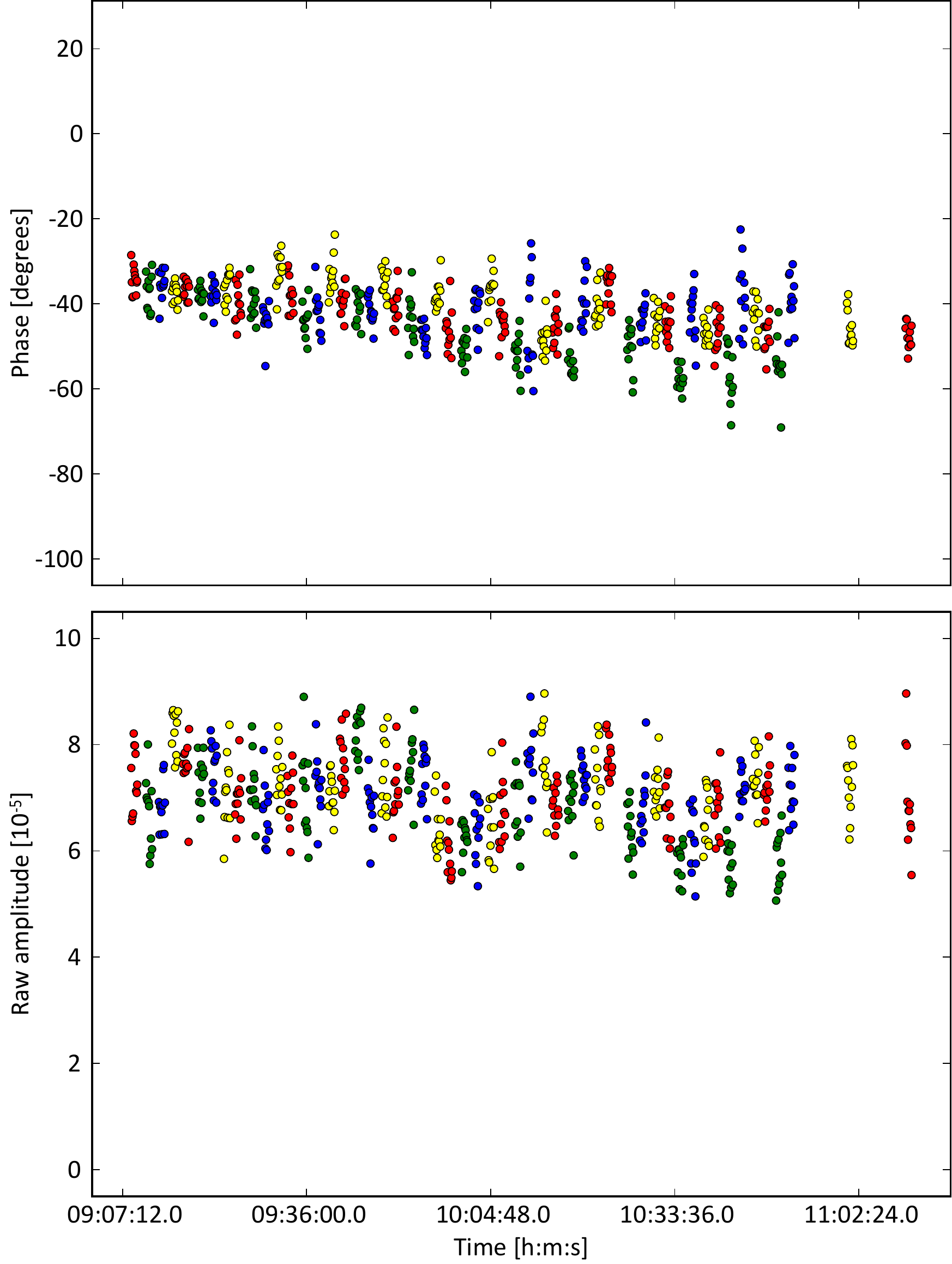}
   \caption{Phase and amplitude of the four-quasar experiment, covering four
            neighboring quasars. The phase transfer as well as the return to
            phase and amplitude are well demonstrated.}
   \label{fig:4quasar}
\end{figure}

Figure~\ref{fig:4quasar} shows the phases (top) and amplitudes (bottom) of all
four sources over the two-hour track binned to 5 seconds timebins on a baseline
of 90\,m. Coherence on all quasars is excellent. Overall, phase RMS ranges from
10 to 25~degrees, while the RMS of the amplitudes is 10 to 20~percent, relative
to the mean value of each source. The observed spread in RMS depends on the
individual source brightnesses and the elevation of each scan. Note that the
uncalibrated raw amplitudes are plotted. No system temperature measurements
were done during this track, preventing us from calibrating the amplitudes.
Similarly, the phase RMS of the third source (in blue) increases as a function
of time due to the decreasing elevation. It can be concluded that the Band~9
performance is excellent in these on-sky tests.

\subsection{Science}

\subsubsection{Science verification: IRAS~16293-2422}

\begin{figure*}[ht!] 
   \centering
   \includegraphics[width=0.85\textwidth]{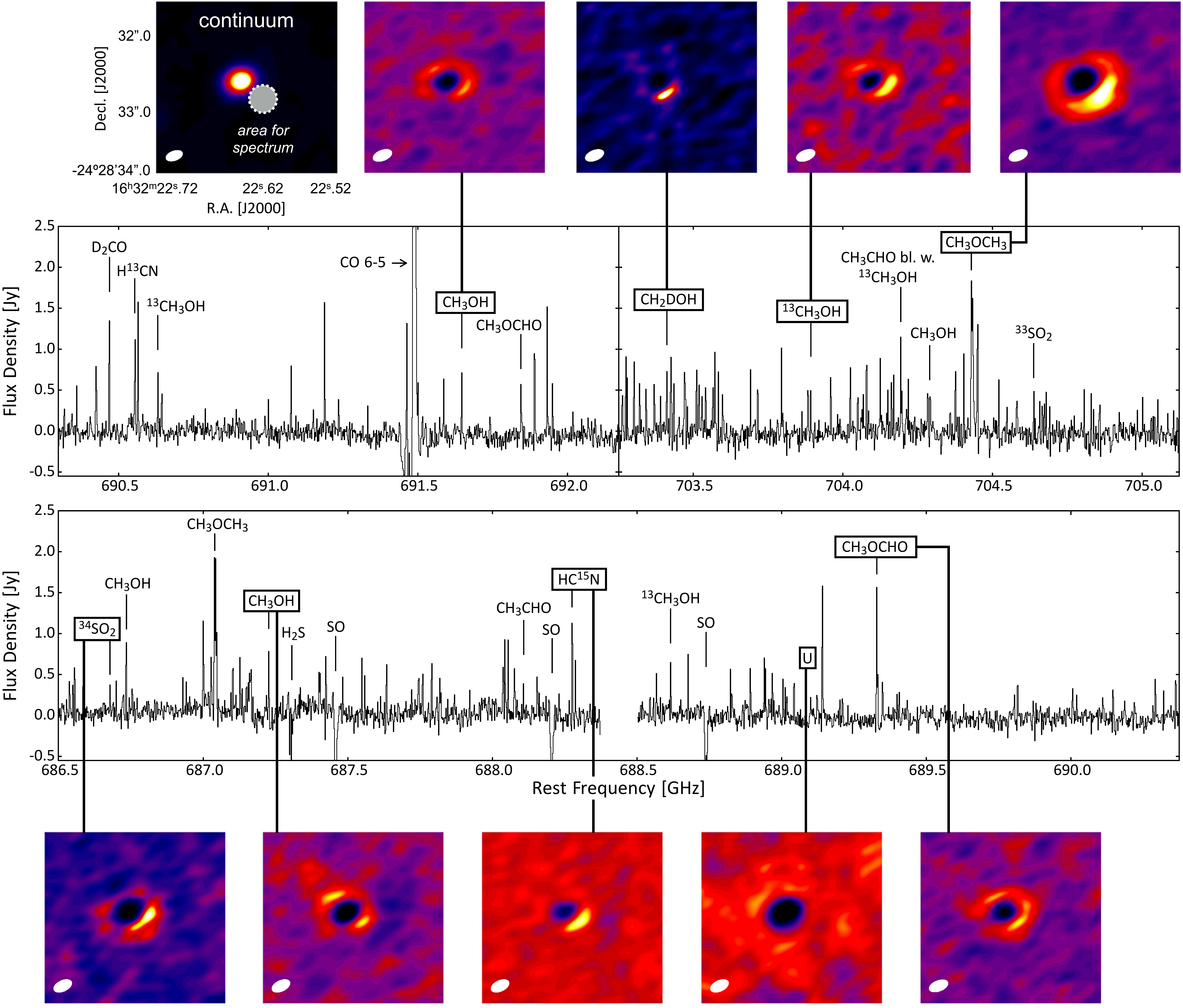}
   \caption{Spectral response of the full observed bandwidth of
            IRAS~16293-2422~B over a region of $0\,\farcs3\times0\,\farcs3$.
            The spectrum has been rebinned to 1.95\,MHz
            ($\sim$0.8\,km\,s$^{-1}$). Known transition identified in Orion~KL
            by \citet{2001ApJS..132..281S} are labeled. Images of the continuum
            and selected lines demonstrate the imaging quality achieved by ALMA
            Band~9.}
   \label{fig:I16293spec}
\end{figure*}

Science Verification of Band~9 was undertaken on April~16 and 17, 2012,
targeting the well-known Class~0 proto-binary
IRAS~16923-2322\footnote{ADS/JAO.ALMA\#2011.0.00007.SV}
\citep[e.g.,][]{1989ApJ...337..858W, Mundy92, Chandler05}. This
$5''$-separation system drives two large-scale bipolar outflows
\citep[e.g.,][]{2004ApJ...608..341S}, and both A and B components show large
numbers of rotational lines of complex organic molecules
\citep[e.g.,][]{1995ApJ...447..760V, 2011A&A...534A.100J}. ALMA Band~6 data
reveal infall motions toward source B \citep{2012A&A...544L...7P}.

A total of 9.2 hours were spent on a seven-point mosaic and the necessary
calibration with ALMA in a Cycle~0 extended configuration, resulting in a
resolution of $\approx$0\,\farcs22. The four spectral windows were chosen to
optimize the available transmission and include the strongest galactic emission
line in Band~9, $^{12}$CO $J$=6--5 at 691.47308\,GHz. More detailed information
on the observations and calibration can be found in the CASA
guide\footnote{http://casaguides.nrao.edu/index.php?title=IRAS16293Band9}.

At the time of writing, the scientific results obtained from this single data
set include the confirmation of the detection of glycolaldehyde, a simple sugar
\citep{2012ApJ...757L...4J}, detection of hot H$_2{}^{18}$O and constraints on
the deuteration of $\sim$100\,K water \citep{2013A&A...549L...3P}, and the
interaction of one of the outflows from source A with the surrounding medium as
traced through warm CO gas \citep{2013A&A...549L...6K, 2012MNRAS.tmpL..35L,
2013ApJ...764L..14Z}.

Figure~\ref{fig:I16293spec} shows the complete Band~9 spectrum of IRAS
16293-2422~B, obtained as part of this Science Verification observation. The
four spectral windows are binned to a resolution of 1.95\,MHz
($\sim$0.8\,km\,s$^{-1}$) and the spectrum is averaged over a
$0\,\farcs3\times0\,\farcs3$ region centered on source B. Over 100 emission
lines are clearly visible in the spectrum. A subset of lines, most of them also
seen by \citet{2001ApJS..132..281S} in Orion KL, are labeled. Most lines are
transitions from organic molecules such as methanol and their $^{13}$C or
deuterated isotopologues. Many unlabeled lines are tentatively associated with
species like C$_3$H$_2$, CH$_3$OCH$_3$ and CH$_3$CHO. Such species have been
detected at lower frequencies \citep[e.g.,][]{2004ApJ...617L..69B,
2004ApJ...616L..27K, 2008A&A...488..959B, 2011A&A...534A.100J}. Both H$^{13}$CN
and HC$^{15}$N are clearly seen. few absorption features are also visble,
caused by simple molecules like H$_2$S in cold foreground gas. Note that the CO
$J$=6--5 line profile is heavily affected by the adopted cleaning strategy,
which was optimized for weak spatially unresolved lines in source B. CO
$J$=6--5 emission extends over a large area \citep{2013A&A...549L...6K}, and is
thus not correctly cleaned. In addition, it suffers from filtering due to the
absence of short baselines.

\subsubsection{Scientific performance}

\begin{figure} 
   \centering
   \hspace{-4mm}\includegraphics[width=0.345\textwidth]{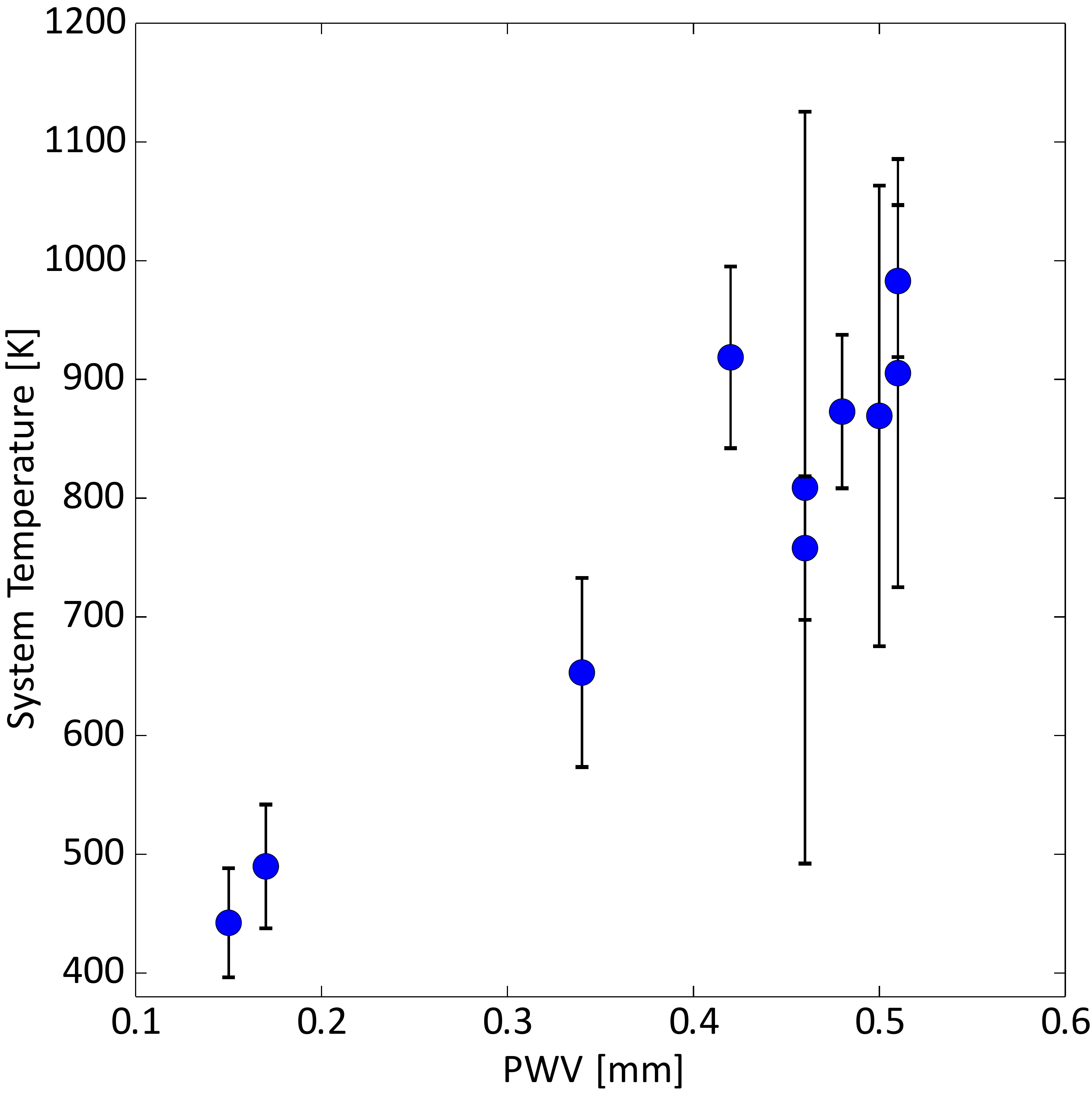}
   \caption{Mean system temperature of antennas for a selection of Cycle~0
            Band~9 science executions.}
   \label{fig:meantsys}
\end{figure}

Thus far, Band~9 has been used under the best conditions available at
Chajnantor. Figure~\ref{fig:meantsys} shows the mean system temperatures,
together with standard deviations, of all unflagged antennes in the various
arrays used for Band~9 scientific executions so far. The atmosphere is the main
noise contributor, but system temperatures smaller than 1000\,K are achieved
for all executions. The executions under the best weather ($<$0.2\,mm of
precipitable water vapor, PWV) approach the mean value expected for the
receiver temperatures of double sideband receivers.

\begin{figure} 
   \centering
   \hspace{-3mm}\includegraphics[width=0.342\textwidth]{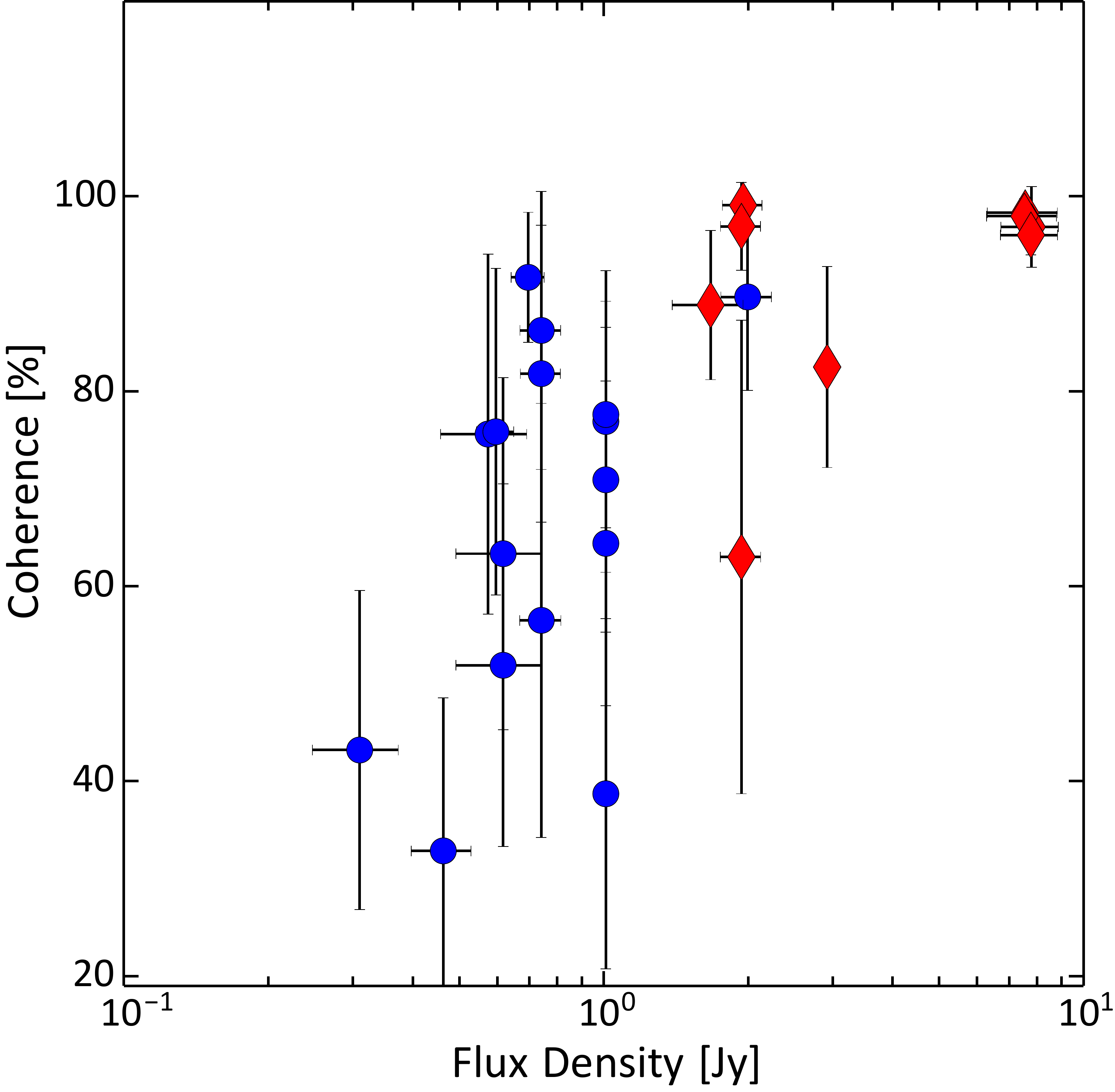}
   \caption{Coherences of bandpass (diamonds) and gain (circles) calibrators
            for Band~9 science executions as a function of flux density.}
   \label{fig:coherence}
\end{figure}

\begin{figure} 
   \centering
   \hspace{-4mm}\includegraphics[width=0.367\textwidth]{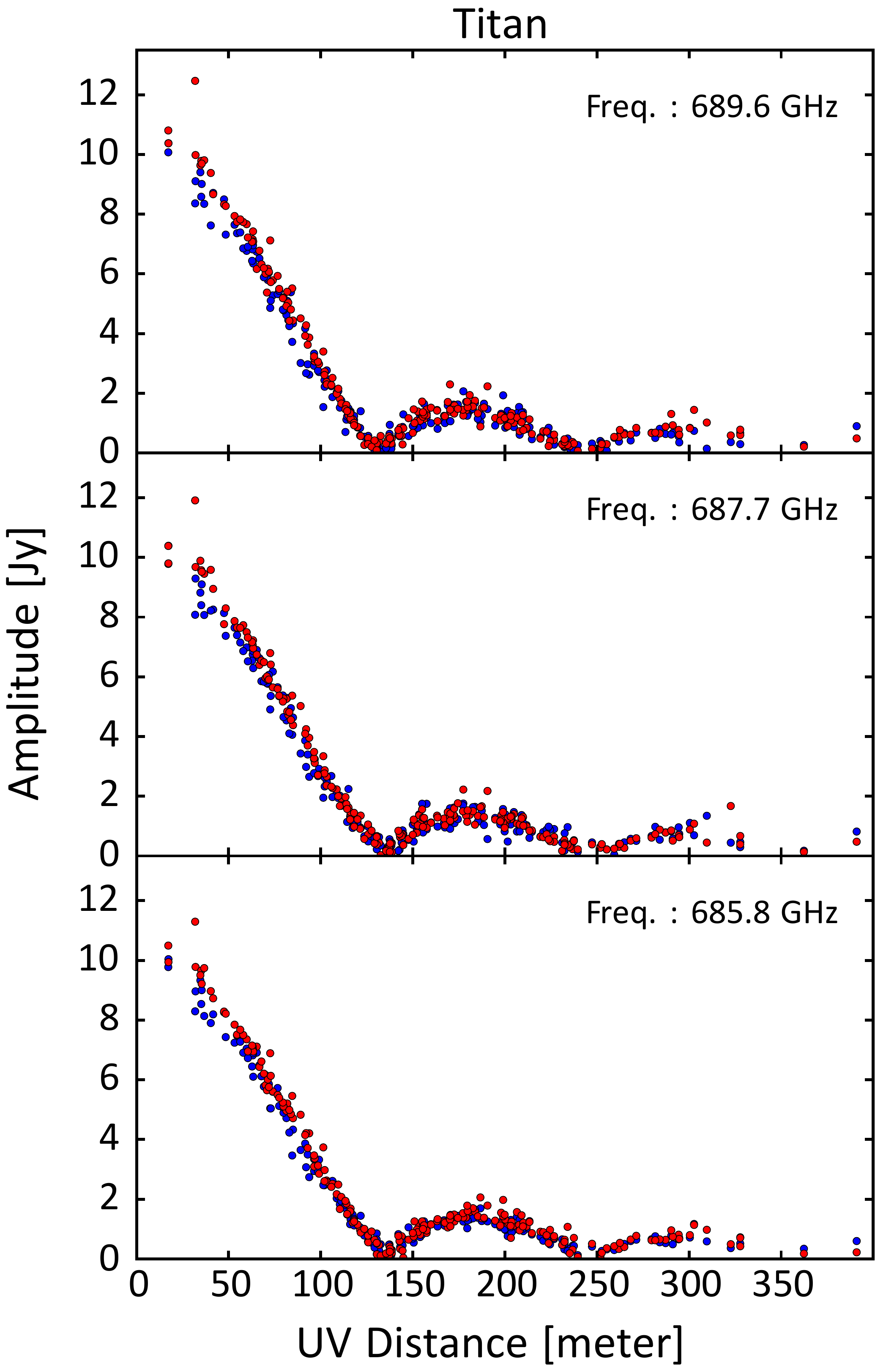} \\
   \vspace{1mm}
   \hspace{-5mm}\includegraphics[width=0.374\textwidth]{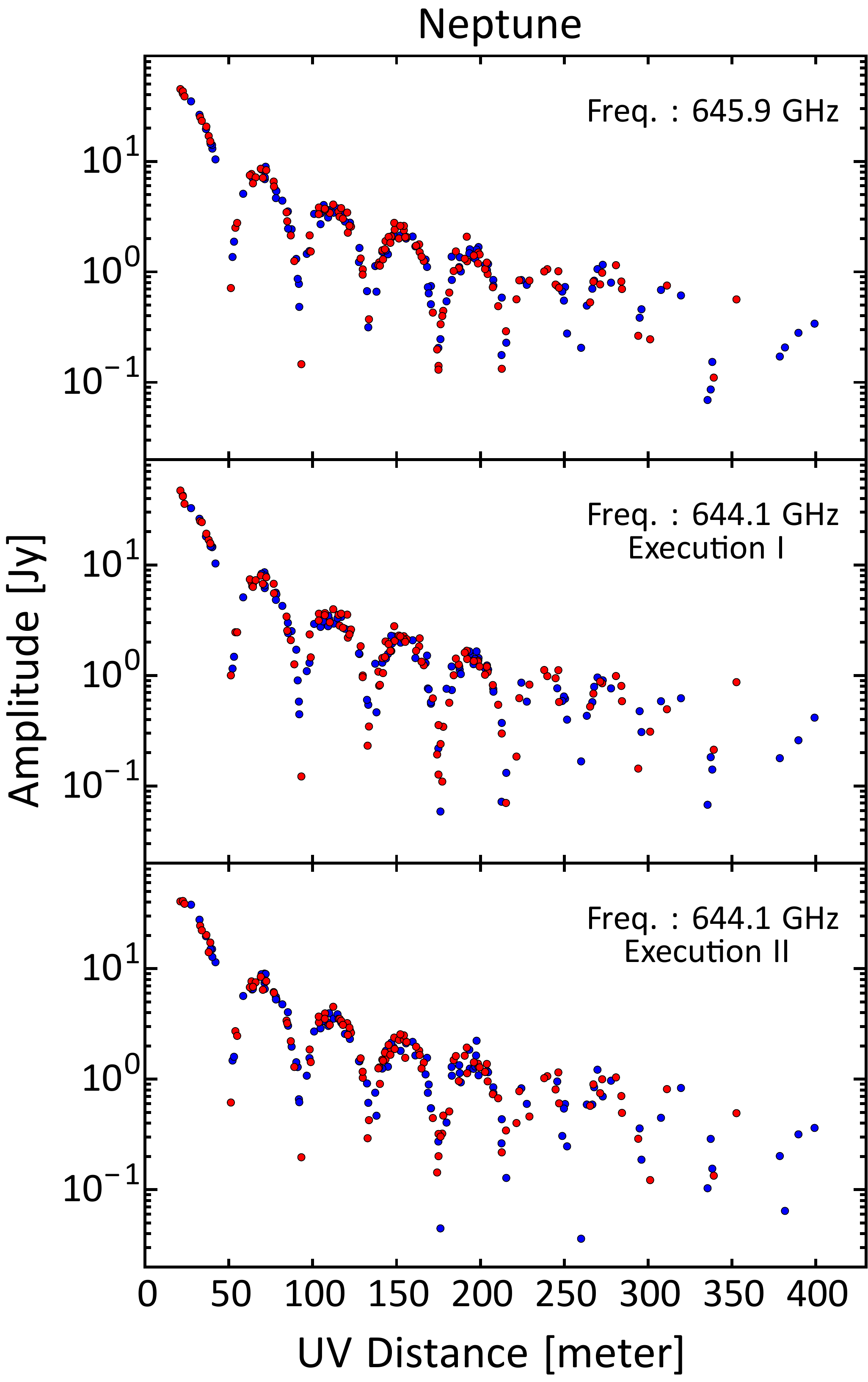}
   \caption{Amplitudes of a sample of flux calibrators (top: Titan, bottom:
            Neptune) as a function of UV-distance. In each case, three spectral
            windows are shown. The plots displaying Neptune data are plotted in
            log(Amplitude) to demonstrate the accuracy of flux calibration in
            resolved objects. Although the amplitudes recover the first eight
            nulls of the resolved Neptune emission, model fits using amplitudes
            up to the fifth null are very accurate. In each plot, blue and red
            points show the different polarizations.}
   \label{fig:fluxcal}
\end{figure}

As a test of scientific quality, Fig.~\ref{fig:coherence} shows the coherence
of all calibrators. Blue circles are the gain calibrators, while red diamonds
are the bandpass calibrators. Coherences are calculated after application of
the water vapor radiometer solutions and plotted for their respective
bandwidths; the full spectral window width for gain calibrators and the
bandpass required channel width for bandpass calibrators. Typical phase
coherences for good bandpass need to be at least 50\% or higher. Note that some
projects required multiple executions within a night, resulting in multiple
measurements of the same calibrator in quick succession. Gain calibration
requires less coherence as all spectral windows can be averaged to obtain
better solutions. The distance of the source to the science target also
influences the quality of the gain solutions; calibrators close to the science
target require less coherence to provide adequate gain solutions.

Figure \ref{fig:fluxcal} shows a few amplitudes of absolute flux calibrators
used for science observations plotted as function of the baseline length. At
Band~9 frequencies, many solar system objects are heavily resolved out (see,
e.g., Neptune). Titan, the Jovian Moons and asteroids are good absolute flux
calibrator candidates in the immediate future of ALMA. However, it is clear
that although ALMA will heavily resolve all solar system objects at baselines
beyond one kilometer, inclusion of a sufficient number of short baselines
allows for accurate absolute flux calibration. Whether this applies to
baselines beyond three kilometers is currently being investigated.

\subsubsection{Early science results using Band 9 observations}

Since the start of ALMA Cycle~0, a number of studies have used observations
from Band~9. \Citet{2013Sci...340.1199V} revealed the existence of a highly
asymmetric dust trap in the disk around the protostar Oph~IRS~48 from Band~9
continuum. Emission lines in the same dataset were used to study the gas
structure within the dust cavity \citep{2014A&A...562A..26B} and to reveal the
existence of warm organic molecules in transition disks
\citep{2014arXiv1402.0392V}. Recently, other Band~9 observations have shown
similar asymmetries in other disks \citep{2014arXiv1402.0832P}.

Another Band~9 project used continuum observations of SN1987A to conclude that
such supernovae may be cosmologically important producers of dust
\citep{2014ApJ...782L...2I, 2014arXiv1409.7811Z}. In the field of evolved
stars, \citet{2013A&A...557L..11B} included observations of $^{12}$CO~6--5 in
their analysis of the gas kinematics of the post-AGB object the Red Rectangle.
Recently, \citet{Decin:2014ub} used Band~9 to reveal the presence of spiral
structure in the inner wind of CW~Leo. Extragalacticly,
\citet{GarciaBurillo:2014gm, 2014arXiv1407.4940V} observed an AGN-driven
outflow in NGC~1068 with the continuum and CO~6--5 at 690\,GHz.


\section{Conclusions}

We have presented details of the design, construction and laboratory testing of
the entire suite of receivers for Band~9 of ALMA. The laboratory results show
that all receivers complied with ALMA specifications prior to integration in
the ALMA antennas. Furthermore, we have presented several examples of the
performance of these receivers as installed in ALMA. Results on phase and
amplitude stability are presented. These results show the extraordinary
performance of the receivers both in the lab and on the telescope. In addition,
Early Science results demonstrate that Band~9 delivers high quality data and
offers exciting prospects for future science observations.

\begin{acknowledgements}

We would like to thank JAO, ESO (especially the Front-End IPT group), NRAO,
NAOJ and the CSV team for their support and collaboration in this project. Part
of this work was supported by NOVA, the Netherlands Research School For
Astronomy. This paper makes use of the following ALMA data:
ADS/JAO.ALMA\#2011.0.00007.SV. ALMA is a partnership of ESO (representing its
member states), NSF (USA) and NINS (Japan), together with NRC (Canada) and NSC
and ASIAA (Taiwan), in cooperation with the Republic of Chile. The Joint ALMA
Observatory (JAO) is operated by ESO, AUI/NRAO and NAOJ.

\end{acknowledgements}

\bibliographystyle{aa}       
\bibliography{B9-DSB_v09.5}

\end{document}